\documentclass[a4paper,fleqn,usenatbib]{mnras}

\usepackage{newtxtext,newtxmath}

\usepackage[T1]{fontenc}
\usepackage{ae,aecompl}

\usepackage{graphicx}	
\usepackage{amsmath}	

\title[Galaxy SED models with PAHs]{Spectral energy distributions of dust and PAHs based on
the evolution of grain size distribution in galaxies}

\author[H. Hirashita, W. Deng and M. S. Murga]{
Hiroyuki Hirashita,$^1$\thanks{E-mail: hirashita@asiaa.sinica.edu.tw}
Weining Deng$^{1,2}$ and
Maria S. Murga$^{3}$
\\
$^{1}$Institute of Astronomy and Astrophysics, Academia Sinica,
Astronomy-Mathematics Building, No.\ 1, Section 4,\\
Roosevelt Road, Taipei 10617, Taiwan \\
$^{2}$Department of Physics, National Taiwan University, Taipei 10617, Taiwan\\
$^3$Institute of Astronomy, Russian Academy of Sciences, Pyatnitskaya str.\ 48, Moscow 119017, Russia}

\date{Accepted XXX. Received YYY; in original form ZZZ}

\pubyear{2020}

\begin{document}
\label{firstpage}
\pagerange{\pageref{firstpage}--\pageref{lastpage}}
\maketitle

\begin{abstract}
Based on a one-zone evolution model of grain size distribution in a galaxy, we calculate
the evolution of infrared spectral energy distribution (SED), considering silicate,
carbonaceous dust, and polycyclic aromatic hydrocarbons (PAHs).
The dense gas fraction ($\eta_\mathrm{dense}$) of the interstellar medium (ISM),
the star formation time-scale ($\tau_\mathrm{SF}$), and the interstellar radiation field intensity
normalized to the Milky Way value ($U$) are the main parameters.
We find that the SED shape generally has weak mid-infrared (MIR) emission in the
early phase of galaxy evolution because the dust abundance is dominated by large grains.
At an intermediate stage ($t\sim 1$ Gyr for $\tau_\mathrm{SF}=5$ Gyr), the MIR emission grows
rapidly because the abundance of small grains increases drastically by the accretion of gas-phase
metals. We also compare our results with observational data of nearby and
high-redshift ($z\sim 2$) galaxies taken by \textit{Spitzer}.
We broadly reproduce the flux ratios in various bands as a function of metallicity.
We find that small $\eta_\mathrm{dense}$ (i.e.\ the ISM dominated by the diffuse phase)
is favoured to reproduce the 8 $\micron$ intensity dominated by PAHs both for the nearby
and the $z\sim 2$ samples.
A long $\tau_\mathrm{SF}$ raises the 8 $\micron$ emission to a
level consistent with the nearby low-metallicity galaxies.
The broad match between the theoretical calculations and the observations supports our
understanding of the grain size distribution, but
the importance of the diffuse ISM for the PAH
emission implies the necessity of spatially resolved treatment for the ISM.
\end{abstract}


\section{Introduction}

Dust grains in the interstellar medium (ISM) absorb the ultraviolet (UV) and optical light and reprocess
it in the infrared (IR). Thus, IR emission from dust is an important tracer of the
star formation activities in galaxies, if the interstellar radiation field (ISRF) is dominated
by young stars \citep[e.g.][]{Buat:1996aa,Inoue:2000aa}.
The star formation activities traced in the IR are important because starbursting
galaxies often emit most of the light in the IR \citep[e.g.][]{Sanders:1996aa} and the contribution from
such IR-luminous population to the total star formation rate in the Universe increases
with redshift, at least up to $z\sim 2$, where $z$ is the redshift
\citep[e.g.][]{Lefloch:2005aa,Perez:2005aa}.
As a consequence, the fraction of star formation activities only traced by dust emission
increases towards high redshift
\citep[e.g.][]{Takeuchi:2005ab,Goto:2010aa,Burgarella:2013aa}.

To understand the nature of IR emission from galaxies, it is crucial to clarify the dust
properties, such as the composition and the grain size distribution. The emissivity in the IR
regime also depends on the dust temperature, which reflects the ISRF intensity
from stellar UV--optical emission \citep[e.g.][]{Li:2001aa}.
There are some detailed modeling efforts for the spectral energy distributions (SEDs) of dust emission
that are well examined or calibrated by the observed SEDs of the Milky Way and nearby galaxies
(\citealt{Desert:1990aa,Dwek:1997aa}; \citealt[][hereafter DL07]{Draine:2007aa};
\citealt{Compiegne:2011aa,Jones:2017aa,Galliano:2018aa}).
Some studies have constructed an SED model in a consistent manner with the stellar spectral
synthesis \citep[e.g.][]{Silva:1998aa,Takagi:2003aa,Noll:2009aa} and some have further calculated the
evolution of SED in galaxy evolution models
\citep[e.g.][]{Granato:2000aa}.
Some recent hydrodynamic simulations of galaxies are processed to calculate the evolution of
dust emission SEDs
by solving radiation transfer under the theoretically predicted spatial distribution of dust
\citep[e.g.][]{Yajima:2015aa,McAlpine:2019aa}.
Analytic modelling of galactic structures is also useful to calculate the emergent SEDs
\citep[e.g.][]{Bianchi:2000ab,Popescu:2011aa}.

Some of the SED models above took into account 
prominent emission features in the mid-IR (MIR), which could give us a clue to
the understanding of grain compositions and grain size distributions.
MIR emission originates mainly from transiently (or stochastically) heated small grains,
whose temperature responds to individual photon energy inputs because of their small heat capacities
and rare encounters with photons \citep{Draine:1985aa}. On the other hand, far-IR (FIR) emission is
usually dominated by
large (mainly sub-micron-sized) grains which are in radiative equilibrium.
The most prominent features in the MIR regime are considered to be caused by
polycyclic aromatic hydrocarbons (PAHs)
\citep{Leger:1984aa,Allamandola:1985aa,Tielens:2008aa,Li:2012aa},
although there are other possible material
candidates such as hydrogenated amorphous carbons (HACs; \citealt{Duley:1993aa}),
quenched carbonaceous composite \citep{Sakata:1984aa}, and
mixed aromatic/aliphatic organic nanoparticles \citep{Kwok:2011aa}.
In this paper, we refer to these MIR features as the PAH features, and assume that they originate from
aromatic structures of carbonaceous materials (not necessarily in exact PAH forms).
Silicate grains also show spectral features at 9.7~$\micron$ and 18 $\micron$, corresponding to
the Si--O stretching and O--Si--O bending mode, respectively \citep{Evans:1994aa}.
The above features give us clues to the grain compositions, which would be difficult to infer from
a featureless SED at FIR wavelengths.\footnote{The MIR and FIR wavelength ranges have
no rigid definition, but they are roughly 3--60 $\micron$ and 60--300 $\micron$, respectively, in
this paper.}

Among various features, prominent PAH features could be used to trace various aspects of galaxy
evolution. Some studies suggest that PAH emission is a useful indicator of the star formation rate
\citep{Forster:2004aa,Peeters:2004aa}.
PAHs also potentially trace the enrichment of carbonaceous materials.
The strong metallicity dependence of the PAH feature strength
\citep[e.g.][]{Engelbracht:2005aa,Draine:2007ab,Galliano:2008aa,Hunt:2010aa,Ciesla:2014aa}
indicates a strong link between the enrichment of PAHs and the evolution of galaxies.
Destruction in supernova (SN) shocks \citep{OHalloran:2006aa} and photo-destruction by
UV irradiation
\citep{Madden:2000aa,Plante:2002aa,Madden:2006aa,Wu:2006aa,Hunt:2010aa,Khramtsova:2013aa}
could be important to produce the metallicity dependence, but the microprocesses and efficiencies
of PAH
destruction in low-metallicity environments are still being debated \citep{Sandstrom:2012aa}.

Some theoretical models succeeded in explaining the strong metallicity dependence of
PAH abundance on metallicity using macroscopic dust enrichment scenarios.
\citet{Galliano:2008aa}
suggested that typical stellar ages of low-metallicity galaxies are too young for
low-mass stars to evolve into asymptotic giant branch (AGB) stars, which could be the source of
PAHs \citep[see also][]{Bekki:2013aa}. However, low-metallicity galaxies may contain
old stellar populations; thus, the metallicity may not be simply related to the galaxy age
\citep[e.g.][]{Kunth:2000aa}.
\citet{Seok:2014aa} modelled the metallicity dependence of PAH abundance by assuming that
the production of small grains by shattering in the ISM is the source of PAHs.
Their model naturally explains the relation between PAH abundance and metallicity because
the shattering efficiency depends strongly on the metallicity (or the dust abundance).
\citet{Rau:2019aa} and \citet[][hereafter HM20]{Hirashita:2020aa} directly calculated the evolution
of grain size distribution and the transition from aliphatic to aromatic carbonaceous grains
by post-processing hydrodynamic simulation data in \citet{Aoyama:2019aa}
\citep[see also][]{Hou:2017aa} and by using a one-zone model, respectively.
They regarded small aromatic grains as PAHs.
As a consequence of their modelling, they successfully explained the relation
between PAH abundance and metallicity
in nearby galaxies. They also reproduced the observed non-linear metallicity dependence of PAH abundance
because of the strong metallicity dependence of small-grain production by
shattering and accretion in their model.
HM20 further predicted extinction curves and showed that it is possible to reproduce the
Milky Way extinction curve by assuming the aromatic grains to be the carriers of the 2175 \AA\ bump.
In HM20, the grain size distribution tends to converge to a shape similar to the
\citet[][MRN]{Mathis:1977aa} distribution, which is known to reproduce the the Milky Way
extinction curve,
and the fraction of carbonaceous dust increases in later epochs because of carbon production by
AGB stars.

The above models including the most recent one of ours mainly focused on
the abundances of dust and PAHs.
Now it is interesting to extend the models to predict dust/PAH emission SEDs, which are directly observed.
Indeed, SEDs could be used to test the dust evolution models that incorporate
grain size distributions and dust compositions.
Emissions at different wavelengths are dominated by
different kinds of dust \citep[e.g.][]{Li:2001aa,Compiegne:2011aa}.
Large and small grains\footnote{The rigid boundary between large and small grains is difficult to
define, but it is roughly around grain radius $a\sim 0.03~\micron$ in this paper.}
contribute mainly to the FIR and MIR emission, respectively, as mentioned above.
The dependence of SEDs on the grain radius and the grain composition is suitable for testing our
dust evolution model (especially HM20).
The usefulness of observed SEDs in constraining the
evolution of grain size distribution is demonstrated by \citet{Relano:2020aa} using the simulation
results in \citet{Hou:2017aa,Hou:2019aa}.
Thus, the main goal of this paper is to develop the dust emission
SED model based on the grain size distributions given by HM20.
This step will serve to test our theoretical understanding of dust
evolution against the observed emission properties.

The SED model developed in this paper could also be useful to extract information on the
evolution of galaxy and dust/PAHs. In particular, we may be able to apply the model to
high-redshift PAH emission to obtain a clue to galaxy evolution. Some IR space telescopes
such as \textit{Spitzer} and \textit{AKARI} have photometric bands sensitive to
PAH emission at various redshifts up to $z\sim 2$ \citep[e.g.][]{Shivaei:2017aa,Kim:2019aa}.
We could also predict the expected PAH
emission luminosity at higher redshifts that is accessible by the current facilities using
the model developed in this paper.

This paper is organized as follows.
In Section~\ref{sec:model}, we describe the SED model for dust emission
based on the dust evolution calculations in MH20.
In Section~\ref{sec:result}, we show the results, which
are compared with nearby galaxy data.
In Section \ref{sec:discussion}, we provide some extended discussions, especially
regarding the PAH optical
properties and the application to high-redshift galaxies.
In Section \ref{sec:conclusion}, we give the conclusion of this paper.
We focus on the IR wavelength range, so that we simply refer to the IR SED as the SED.

\section{Model}\label{sec:model}

We calculate the SEDs of dust and PAHs based on the galaxy evolution model developed in our
previous paper (HM20). The model predicts the grain size distribution and the grain
composition (silicate, aromatic carbon, and non-aromatic carbon) in a manner consistent
with the chemical enrichment and dust processing in the galaxy.
We utilize the grain size distribution of each composition to calculate
the emission SED using \citet{Draine:2001aa}'s method.
We review these frameworks and explain the key physical parameters.

\subsection{Chemical enrichment}\label{subsec:chem}

The enrichment of the galaxy with metals is treated using a chemical evolution model.
We treat the galaxy as a one-zone closed box.
We adopt the star formation rate as a function of time $t$ as $\psi (t)\propto\exp (-t/\tau_\mathrm{SF})$,
where we give the decaying time-scale of star formation rate,
$\tau_\mathrm{SF}$ (referred to as the star formation time-scale), as a free parameter.
Based on this model, we
calculate the metal and dust enrichment by stars (SNe and AGB stars).
We also compute the metallicity and the silicon and
carbon abundances. The metal and dust yields in the literature are adopted
(see HM20 for the references). We also calculate the SN rate (denoted as $\gamma$) by
assuming that stars in a mass range of 8--40 M$_{\sun}$ end their lives as SNe.
We adopt the Chabrier initial mass function \citep{Chabrier:2003aa} 
with a stellar mass range of 0.1--100 M$_{\sun}$.

\subsection{Grain size distribution of various components}\label{subsec:gsd}

The HM20 model computes the evolution of grain size distribution as
formulated by \citet{Hirashita:2019aa}. The formulation was based on
\citet{Asano:2013aa} and \citet{Nozawa:2015aa}.
It also calculates the evolution of the fraction of
various dust components (silicate, aromatic carbon,
and non-aromatic carbon).
We only review the outline and refer the interested reader to HM20 for further details.

The grain size distribution as a function of grain radius $a$ is denoted as $n_i(a)$
($i$ distinguishes the dust compositions),
and is defined such that $n_i(a)\,\mathrm{d}a$ is the number density of grains
with radii between $a$ and $a+\mathrm{d}a$. The grain mass, $m$, is related to $a$
as $m=(4\upi /3)a^3s$, where $s$ is the grain material density
(we assume that $s=2.24$ g cm$^{-3}$, which is appropriate for graphite,
in the calculation of grain size distribution.).
{To avoid the complexity arising from the collisions among different dust species,
we calculate the evolution of grain size distribution without distinguishing the species.
Here we represent the dust properties by those of graphite
following HM20. HM20 also showed that
adopting silicate properties instead does not alter the resulting evolutionary behaviour of grain size distributions
except for some detailed changes in the efficiencies of shattering and coagulation.
These details are not important for our conclusions. We note that, for the purpose of calculating the SED,
we do consider the difference in the grain species by decomposing the resulting grain size distributions
into multiple components as explained later.}

We calculate the evolution of grain size distribution
by considering the following processes: dust production by stars (SNe and AGB stars),
dust destruction by SN shocks sweeping the ISM, dust growth by the accretion of gas-phase metals in
the dense ISM,
grain growth by coagulation (grain--grain sticking) in the dense ISM and grain disruption by
shattering (fragmentation) in the diffuse ISM.
The dust enrichment by stars is calculated in the chemical evolution model described
in Section \ref{subsec:chem}, and the produced dust by stellar sources is
distributed in each grain radius bin by assuming the lognormal grain size distribution
centred at 0.1 $\micron$ with a standard deviation of 0.47 \citep{Asano:2013aa}.

Since accretion and coagulation occur only in the dense ISM and shattering takes place
only in the diffuse ISM, we need to specify the mass fraction of the dense ISM,
$\eta_\mathrm{dense}$ (the diffuse ISM occupies the rest, $1-\eta_\mathrm{dense}$).
We assume that the diffuse and dense ISM components have
$(n_\mathrm{H}/\mathrm{cm}^{-3},\, T_\mathrm{gas}/\mathrm{K})=(0.3,\, 10^4)$
and $(300,\, 25)$, respectively,
where $n_\mathrm{H}$ is the hydrogen number density and $T_\mathrm{gas}$ is the gas temperature.
{It is likely that $\eta_\mathrm{dense}$ changes as the galaxy evolves. 
Indeed, there is an indication that the relative weight of the dust emission
from the dense and diffuse media changes with star formation activities (or galaxy evolution)
\citep{da-Cunha:2010aa}. Since
what regulates the evolution of $\eta_\mathrm{dense}$ is not fully understood, we simply treat
$\eta_\mathrm{dense}$ as a constant free parameter following HM20. This simple treatment is
useful in making the effect of $\eta_\mathrm{dense}$ easy to interpret.}

{As mentioned above, the calculated grain size distribution}
does not distinguish the grain compositions but
represent the total grain size distribution. Thus, we separate the grain size distribution into
silicate and carbonaceous grains in a manner consistent with the abundance ratio of
Si to C at each age. {This is necessary to calculate the SED, which depends strongly on the
grain species.} The mass fraction of silicate to the total dust is denoted as $f_\mathrm{sil}$,
and we assume that the material density for silicate is $s=3.5$ g cm$^{-3}$.
We further divide the carbonaceous component into aromatic and non-aromatic species.
In our model, small aromatic carbonaceous grains (typically $a\sim 3$--50~\AA) represent PAHs
(note that that 3 \AA\ is the minimum radius of the grains assumed in the model).
We define the aromatic fraction, $f_\mathrm{ar}(a,\, t)$, as the fraction of
the aromatic carbonaceous grains to the
total carbonaceous grains at each grain radius $a$.
We assume that aromatization and aliphatization, whose rates are given in HM20
\citep[based on][]{Murga:2019aa},
occur in the diffuse and dense ISM, respectively
(we include aromatization by UV irradiation and SN shocks, and aliphatization by
hydrogen attachment and metal accretion).
As shown by HM20, $f_\mathrm{ar}\simeq 1-\eta_\mathrm{dense}$ in most of the relevant grain radius
range as a result of equilibrium between aromatization and aliphatization, both of which occur on
a time-scale much shorter than that of ISM phase exchange.
Since the equilibrium value of $f_\mathrm{ar}$ is insensitive
to the ISRF, we simply use the ISRF appropriate for
the Milky Way \citep{Mathis:1983aa} to calculate $f_\mathrm{ar}$.
We neglect photo-destruction of PAHs and concentrate on
the processes included above for the evolution of grain
size distribution (i.e.\ PAHs are simply treated as small aromatic grains in our model).
As discussed by HM20, the diffuse ISRF included in our one-zone model
is not likely to reduce the PAH abundance by
photodestruction faster than the continuous supply of PAHs by shattering.
We are able to give some quantitative
estimates based on \citet{Allain:1996aa} and \citet{Murga:2019aa}.
{We performed the following experimental calculation.
We basically adopted the ISRF spectrum from \citet{Mathis:1983aa} and scaled it with the
intensity $U$ ($U=1$ corresponds tot he Milky Way ISRF in the solar neighbourhood).
To realize a hard UV spectrum dominated by massive stars in dwarf galaxies,
we replaced the UV spectrum at wavelengths 912--2500 \AA\ of \citet{Mathis:1983aa} with
a hard spectrum $\nu I_\nu\propto\nu^2$, which roughly represents young
($<10^7$ yr) and low-metallicity ($<0.1$ Z$_{\sun}$)
stellar population \citep{Wilkins:2013aa}, with the continuity at
2500 \AA.\footnote{For the other calculations in this paper,
we simply used the spectral shape of \citet{Mathis:1983aa}. This modification of the spectral shape
is only for the experimental purpose here.} In this case,}
PAHs with $a<5$~\AA\ could be destroyed within the
cosmic age if $U\sim 1$. If we assume $U=100$ appropriate for some low-metallicity
galaxies, the PAH destruction time-scale for $a>4$~\AA\ is longer than
the time-scale on which the supply of small grains by shattering occurs
($10^8$--$10^9$ yr). Thus, only the smallest PAHs ($a=3$--4 \AA) could be depleted by
photodestruction.
This significantly decreases the 3.3 $\micron$ PAH intensity, which we do not focus on in this
paper. PAHs with $a>5$~\AA\
are hard to be destroyed because of the stable carbon skelton,
so that the main role of UV irradiation is dehydrogenation.
Since most of the contribution to the 8 $\micron$ bands comes from PAHs with
$a\gtrsim 5$~\AA\ \citep{Draine:2007aa}, we expect that our results are not significantly affected
by photodestruction.
There could be some paths of PAH destruction by a local strong UV radiation field
(e.g.\ destruction of super-hydrogenated carbons), but such a local and particular process is
difficult to include in our one-zone model.
Therefore, we leave the effect of local strong radiation
and the dependence on detailed chemical structures of carbonaceous dust for a future work.

Overall, a smaller value of $\eta_\mathrm{dense}$ leads to more small grains
(because of more shattering and less coagulation) and
a higher aromatic fraction (because of more aromatization).
Thus, the PAH abundance is more enhanced in the case of smaller $\eta_\mathrm{dense}$.
The PAH abundance also depends on the star formation time-scale $\tau_\mathrm{SF}$.
Note that the silicate fraction is high at young ages,
since silicon, compared with carbon, is preferentially
produced by SNe. Thus, the aromatic features tend to be suppressed for
shorter $\tau_\mathrm{SF}$. At the same metallicity, the PAH abundance is higher in the
case of longer $\tau_\mathrm{SF}$ because of an older age (i.e.\ it takes more time to
reach a certain metallicity if $\tau_\mathrm{SF}$ is longer).

\subsection{SED model}\label{subsec:SED}

We calculate the SED using the framework developed by \citet{Draine:2001aa} and \citet{Li:2001aa}.
The SED depends on the grain composition and the grain size distribution.
The distribution function of grain temperature for each $a$ is calculated by considering the transition of the grain energy state
due to the heating by the ISRF and the cooling by IR radiation.
We adopt the ISRF spectrum from \citet{Mathis:1983aa}
but scale it with a parameter $U$ ($U=1$ corresponds to the Milky Way ISRF in the solar
neighbourhood).
Using the resulting grain temperature distribution of the grain component $i$
(denoted as $\mathrm{d}P_i/\mathrm{d}T$, which is a function of $a$), we calculate the SED
[the intensity of dust emission per hydrogen at frequency $\nu$ (with corresponding
wavelength $\lambda$), denoted as $I_\nu (\lambda )$] by
\begin{align}
I_\nu (\lambda )=\sum_i
\int_0^\infty\mathrm{d}a\frac{1}{n_\mathrm{H}}n_i(a)\upi a^2Q_\mathrm{abs}(a,\,\nu )
\int_0^\infty\mathrm{d}T\, B_\nu (T)
\frac{\mathrm{d}P_i}{\mathrm{d}T},\hspace{-1cm}\nonumber\\
\end{align}
where $Q_\mathrm{abs}(a,\,\nu)$ is the absorption cross-section normalized to the
geometric cross-section ($\upi a^2$) as a function of grain radius $a$ and frequency $\nu$,
and $B_\nu (T)$ is the Planck function at frequency $\nu$ and temperature $T$.
We assume that the IR radiation
is optically thin. For the SEDs, we show the above intensity per hydrogen in this
paper.

We treat $U$ as a free parameter, although it may depend on the star formation activity.
This is because $U$ is also affected by various radiation transfer effects, especially
the shielding of dust.
Such radiation transfer effects cannot be treated by our one-zone model.
We also concentrate on the mean ISRF in the galaxy, neglecting the inhomogeneity in the radiation field
intensity.

For the optical properties of dust, we adopt two models from DL07 and
\citet[][hereafter J13]{Jones:2013aa}, which are often used to model the SEDs of
the Milky Way and nearby galaxies.
The dust optical properties from these papers are referred to as the DL07/J13 model.
The DL07 model is based on the mixture of silicate, graphite, and PAHs.
It is straightforward to use their astronomical silicate properties for the silicate component
in our model. For the aromatic carbon component, we adopt their carbonaceous component,
whose optical properties have a smooth transition from PAHs to graphite around $a\sim 50$ \AA.
We adopt the ionization fraction of PAHs as a function of grain radius following DL07
(originally from \citealt{Li:2001aa}), who calculated the ionization fractions
in various ISM phases and averaged them with an appropriate weight.
For the non-aromatic component, we simply adopt graphite from \citet{Draine:1984aa}
(i.e.\ DL07's carbonaceous properties without PAH features).
According to \citet{Hirashita:2014aa}, the difference in the mass absorption coefficient
between amorphous carbon taken from \citet{Zubko:1996aa}
and graphite above is 40 per cent around the peak of the
IR SED ($\lambda\sim 150~\micron$).
Keeping this uncertainty in mind, we simply stick to the available grain properties in DL07
(and their series of papers).
For the J13 model, we adopt amorphous silicate for
the silicate component, while we apply
hydrogenated aromatic carbon (a-C:H) and
aromatic carbon (a-C) for the non-aromatic and
aromatic components, respectively. More specifically,
we adopt the materials with low ($E_\mathrm{g}=0.1$~eV) and high
($E_\mathrm{g}=2.67$~eV) band gap energy for the aromatic and non-aromatic species,
respectively. {For the ionization fractions of the carbonaceous species, we adopt the
same value as the above PAH component in the DL07 model (i.e.\ the values taken from DL07).}
As pointed out in HM20, our predicted grain size distributions
reproduce the Milky Way
extinction curve better with the DL07 model than with the J13 model.
In this paper, we investigate both dust
properties, keeping in mind that the DL07
dust materials combined with our grain size
distribution reproduce the Milky Way extinction curve
relatively well. As discussed below, the two dust models have some common
results, which could strengthen our conclusions.

\subsection{Observational data for comparison}\label{subsec:obs}

We aim at examining if our model reproduces the observed trend in
SEDs for galaxies at various evolutionary stages.
We use metallicity as an indicator of the galaxy evolutionary stage
as done in many models of chemical/dust evolution models
\citep[e.g.][]{Lisenfeld:1998aa,Dwek:1998aa,Inoue:2003aa,Zhukovska:2008aa}.

We adopt comprehensive nearby galaxy samples that include
the \textit{Spitzer} data. The \textit{Spitzer} bands are suitable for tracing the
emission from various dust components including PAHs.
We take the samples from the following two papers.
One is the union of the KINGFISH (Key Insights on Nearby Galaxies:
A Far-Infrared Survey with \textit{Herschel}) and SINGS
(\textit{Spitzer} Infrared Nearby Galaxies Survey) samples, for which
\citet{Dale:2017aa} compiled data from UV to radio wavelengths.
We adopt the oxygen abundance from \citet{Moustakas:2010aa} for this sample by adopting
the \citet{Kobulnicky:2004aa} metallicity calibration in their table 9.
Among the 79 galaxies in their sample, we adopt 75 galaxies excluding NGC~3034 [M82;
because of the saturation in the \textit{Spitzer} MIPS
(Multiband Imaging Photometer) bands],
DDO 154, Holmberg IX, and M81 Dwarf B (because of no detection in the
relevant \textit{Spitzer} bands).
The other is the sample of nearby star-forming galaxies with a large metallicity
range, taken from \citet{Engelbracht:2008aa}.
We exclude UM 420 because
of no available MIPS data. Their compiled data also include the oxygen abundance.
Because of the crowded data points in the diagrams to be shown below, we do not explicitly
show error bars. The typical errors in the luminosity and the metallicity are 10 per cent
and 0.1 dex, respectively.
We always show the metallicity normalized to the solar value: For the observational data,
we assume the metallicity to scale with the oxygen abundance and adopt
$12+\log (\mathrm{O/H})_{\sun}=8.76$ for the solar oxygen abundance and $Z_{\sun}=0.015$ for the solar metallicity
\citep{Lodders:2003aa}.
\citet{Ciesla:2014aa} also compiled the data for the HRS (\textit{Herschel} Reference Survey)
data. Their PAH abundances roughly overlap
($12+\log (\mathrm{O/H})>8.1$) with those of the SINGS sample at high metallicities.
Because of their smaller
coverage in metallicity (and the metallicity dependence is critical in this paper),
we do not adopt their data, keeping in mind that they
also show a similar correlation between PAH abundance
and metallicity.

To constrain our model, we use the \textit{Spitzer} IRAC
(Infrared Array Camera) 8 $\micron$, and
MIPS 24, 70 and 160 $\micron$ bands.
The two longest wavelengths are used to constrain the SED of large grains which are
in radiative equilibrium with the ISRF. In other words, the 70 and 160 $\micron$ bands can be used to
constrain $U$. The 8 and 24 $\micron$ bands are dominated by small grains that are not in
radiative equilibrium (or are stochastically heated).
The 8 $\micron$ emission is dominated by PAH emission if the PAH abundance is high enough, and
thus can be used as an indicator of PAH emission strength \citep{Engelbracht:2005aa}.
The 24 $\micron$ band is not affected by any particular PAH feature, but it reflects the overall
small-grain population. Thus, the set of these \textit{Spitzer} bands is suitable for the
purpose of investigating the grain size distribution and the PAH abundance.

The total IR (TIR) luminosity of the observational sample is evaluated using the
8, 24, 70, and 160 $\micron$ luminosities\footnote{More precisely, the central wavelengths are
7.9, 24, 71, and 160 $\micron$ (DL07), but we simply describe the wavelength
as written in the text.} as (DL07)
\begin{align}
L_\mathrm{TIR}\approx 0.95\langle\nu L_{\nu }\rangle_{8}+
1.15\langle\nu L_{\nu }\rangle_{24}+\langle\nu L_{\nu }\rangle_{70}+
\langle\nu L_{\nu }\rangle_{160}\, ,\label{eq:TIR}
\end{align}
where the luminosity density weighted for the filter band pass
is multiplied by the central frequency, and is denoted as $\langle\nu L_\nu\rangle_{\lambda_\mathrm{c}}$,
($\lambda_\mathrm{c}$ is the central wavelength of the band in units of
$\micron$).
For the \citet{Dale:2017aa} sample, we subtract the contribution from stellar emission estimated
using the IRAC 3.6 $\micron$ data (DL07):
\begin{align}
&L^\mathrm{ns}_{\nu }(8~\micron ) = L_{\nu }(8~\micron ) -
0.232 L_{\nu }(3.6~\micron ),\\
&L^\mathrm{ns}_{\nu }(24\mu m) = L_{\nu }(24~\micron ) - 0.032
L_{\nu }(3.6~\micron ).
\end{align}
For the \citet{Engelbracht:2008aa} sample, we use their stellar-subtracted fluxes.
Because there is no risk of
confusion, we hereafter omit the superscript `ns'.
For the theoretical SEDs, we also evaluate the total intensity, $I_\mathrm{TIR}$, using
equation (\ref{eq:TIR}) but replacing the luminosities with the intensities.
The filter response of the appropriate band is considered for the theoretically derived intensities.

Our SED model outputs the intensity (per hydrogen), while the observational data show the total
luminosity.
We  denote the intensities weighted for the band pass, omitting the bracket
$\langle~\rangle$ but clarifying the wavelength,
as $I_\nu (\lambda )$.
We mainly compare the ratios; for example, $I_\nu (70~\micron )/I_\nu (160~\micron )$
in our model is compared with the observed luminosity ratio between 70 $\micron$ and 160 $\micron$.
It is sometimes convenient to multiply $\nu$ corresponding to the
central wavelength [denoted as $\nu I_\nu (\lambda )$] to make the physical dimension
identical to $I_\mathrm{TIR}$.

For the basic presentations of SED evolution in Sections \ref{subsec:SED_result}
and \ref{subsec:J13},
we also plot the observed Milky Way SED. We take the data for the diffuse high Galactic latitude
medium from \citet{Compiegne:2011aa}, which can be used as typical dust emission data in the
solar neighbourhood \citep[see also][]{Desert:1990aa}. We adopt their data taken by the
\textit{COBE}/DIRBE, \textit{Herschel}/PACS, \textit{Herschel}/SPIRE, and \textit{Planck}/HFI
at wavelengths shorter than 1 mm. Since we already know that $U\sim 1$ is appropriate for
the Milky Way, it is not necessary to adjust $U$, while for other galaxies, $U$ is unknown.
Thus, we only adopt the Milky Way for
a direct comparison with the SEDs.

For PAHs, spectroscopic data could also be useful; however, comparison with
spectroscopic data is too sensitive to assumed optical properties and
ionization states of PAHs.
As we will show below, comparison with photometric data already
puts meaningful constraints on the model; thus, the purpose of this paper is specifically to
examine if the broad SED shapes traced by photometric data could be reproduced using our models.
We leave detailed comparison with spectroscopic data \citep[e.g.][]{Hunt:2010aa} for a future work.

\section{Results}\label{sec:result}

\subsection{Grain size distribution}\label{subsec:size}

We first review the resulting grain size distributions in HM20.
The interested reader is referred to
HM20 for the detailed discussions on the evolution of grain size distributions.


The grain size distribution is dominated by large ($a\sim 0.1$--1~$\micron$) silicate grains
in the early epoch of galaxy evolution when the dust formation is dominated by stellar sources.
In the Milky Way-like star formation history with $\tau_\mathrm{SF}=5$ Gyr, this phase continues up to
$t\sim 0.3$ Gyr.
After this age, the grain size distribution is modified by interstellar processing,
which is sensitive to the assumed dense gas fraction, $\eta_\mathrm{dense}$.
The evolution in the `fiducial' case with $\eta_\mathrm{dense}=0.5$ and $\tau_\mathrm{SF}=5$ Gyr is
described as follows.
Shattering gradually produces small grains at $t\sim 0.3$--1 Gyr.
At $t\sim 1$ Gyr, the metallicity becomes high enough for accretion to increase the
abundance of small ($a\sim 0.001$--0.01~$\micron$) grains
(note that accretion is efficient for small grains since they have high
surface-to-volume ratios).
After this phase of efficient accretion, coagulation starts to have a large influence on
the grain size distribution by converting small grains to large grains.
In this phase, the grain size distribution becomes similar to the MRN grain size distribution
($\propto a^{-3.5}$).

The evolution of grain size distribution depends strongly on the dense gas fraction, $\eta_\mathrm{dense}$.
For small $\eta_\mathrm{dense}(\sim 0.1)$, coagulation does not occur efficiently while shattering is efficient, so the grain size
distribution at $t\gtrsim 1$~Gyr is dominated by small ($a\lesssim 0.03~\micron$) grains.
In contrast, for large $\eta_\mathrm{dense}(\sim 0.9)$, grains as large as $a\sim 1~\micron$
form efficiently by coagulation at $t\gtrsim 3$ Gyr.

For $\tau_\mathrm{SF}\sim 5$ Gyr,
the grain composition is dominated by silicate with $f_\mathrm{sil}\sim 0.9$ at $t<1$ Gyr and
decreases down to $f_\mathrm{sil}\sim 0.7$ at later epochs.
The aromatic fraction, $f_\mathrm{ar}$, is broadly equal
to $1-\eta_\mathrm{dense}$ as mentioned in
Section \ref{subsec:gsd}. Moreover, if $\eta_\mathrm{dense}$ is smaller, grains are more dominated
by small grains as mentioned above, which leads to a higher PAH abundance.
As a consequence, the PAH abundance is affected by
the age and $\eta_\mathrm{dense}$ in such a way that a system with an old age and
a large fraction of the diffuse ISM tends to have a high PAH abundance.

The evolution of dust and PAHs is also affected by $\tau_\mathrm{SF}$.
A similar grain size distribution is obtained for the same value of $t/\tau_\mathrm{SF}^{1/2}$
as shown by HM20. At a certain fixed metallicity, the
case with shorter $\tau_\mathrm{SF}$ has a higher silicate fraction and a lower
small-grain abundance simply because of a younger age
(thus, less AGB stars and less time for dust processing).
This means that the PAH abundance at a certain
metallicity is suppressed for short $\tau_\mathrm{SF}$.

The above results already shown by HM20 give a basis on which we interpret the SED evolution
described in the following subsections.

\subsection{Evolution of SED}\label{subsec:SED_result}

\begin{figure}
\includegraphics[width=8cm]{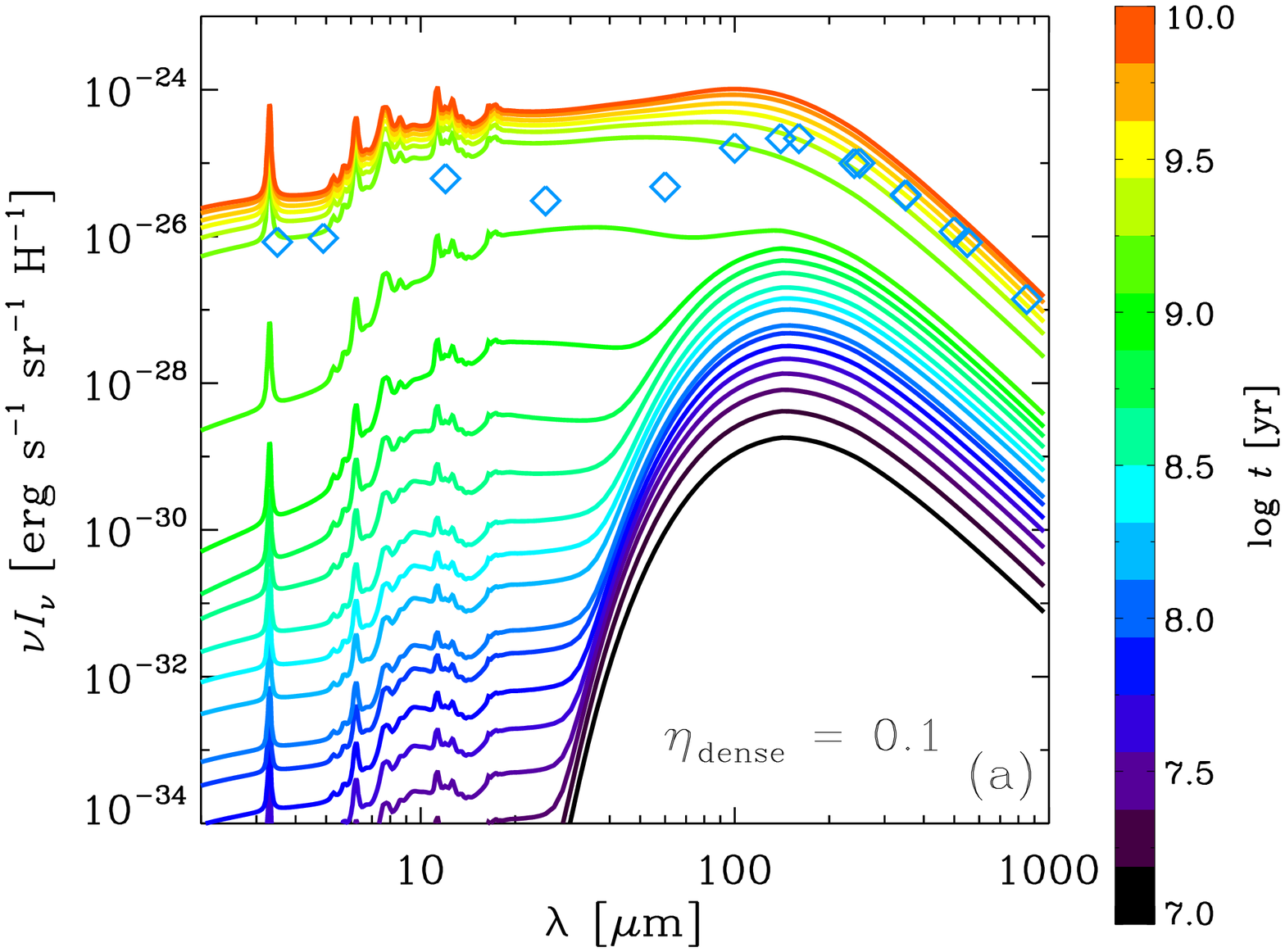}\\
\includegraphics[width=8cm]{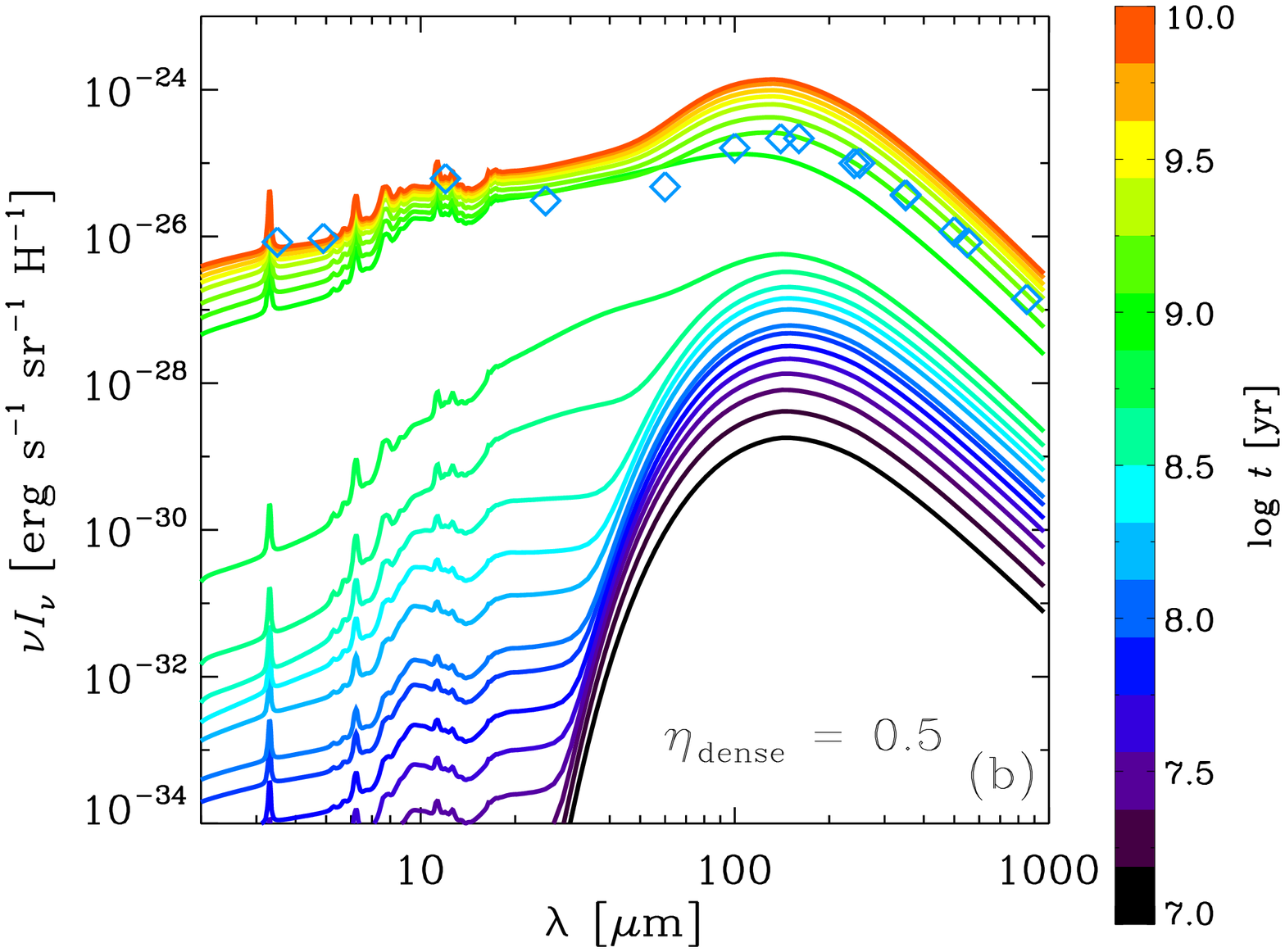}\\
\includegraphics[width=8cm]{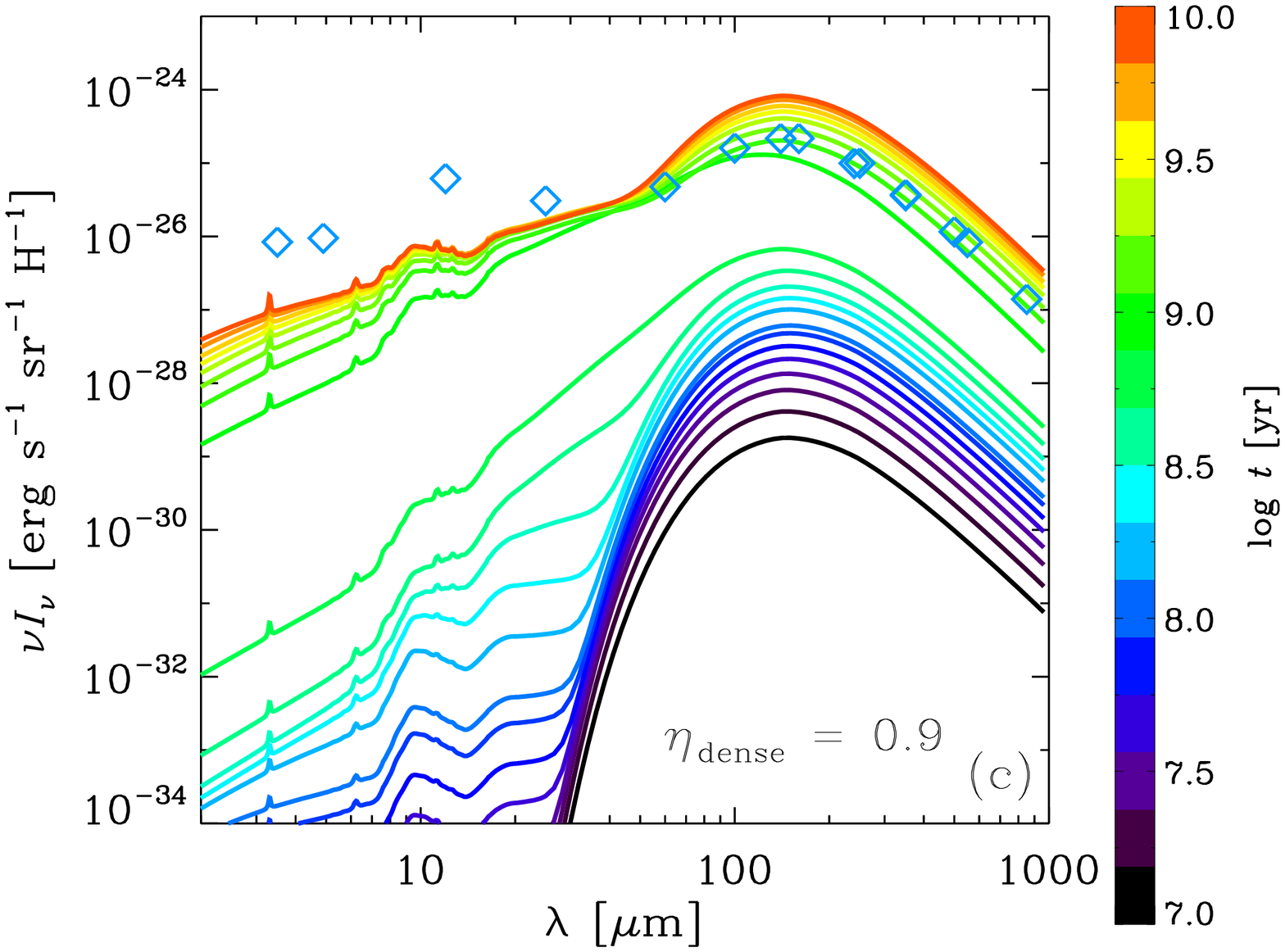}
\caption{SEDs with $\tau_\mathrm{SF}=5$ Gyr and $U=1$
from $t=10^7$ to $10^{10}$ yr (with a logarithmically equal interval) under the DL07
grain compositions.
The colour indicates the age as shown in the colour bar.
Panels (a), (b), and (c) show the results for $\eta_\mathrm{dense}=0.1$, 0.5, and 0.9, respectively.
We also show the observational data for the Milky Way from \citet{Compiegne:2011aa}
(taken by the \textit{COBE}/DIRBE, \textit{Herschel}/PACS, \textit{Herschel}/SPIRE, and
\textit{Planck}/HFI at wavelengths shorter than 1 mm; Section \ref{subsec:obs}).
The errors of these data are smaller than the symbol size.
\label{fig:sed}}
\end{figure}

As explained above, the grain size distribution and the dust compositions are affected by 
$\tau_\mathrm{SF}$ and $\eta_\mathrm{dense}$; thus, the SEDs also depend on these parameters.
In addition, the SEDs also rely on the ISRF intensity, $U$.
In this section, we focus on the case with $\tau_\mathrm{SF}=5$ Gyr, which is appropriate for
Milky Way-like galaxies. 
We also adopt $U=1$, the typical ISRF intensity in the solar neighbourhood of the Milky Way.
We examine various values of $\eta_\mathrm{dense}$ ($=0.1$, 0.5, and 0.9).
We first concentrate on the results for the DL07 grain composition model
(we always adopt the DL07 model unless otherwise stated)
and discuss the J13 model in Section \ref{subsec:J13}.

We show the evolution of SED in Fig.\ \ref{fig:sed}
between $t=10^7$ and $10^{10}$ yr.
The intensity increases almost monotonically with age because of dust enrichment.
We observe a big jump in the overall dust emission around
$t\sim 1$ Gyr, which corresponds to the epoch when dust growth by accretion drastically increases
the dust abundance. Since the small-grain abundance particularly grows,
the increase in the MIR emission (including PAH emission) is more prominent than that in the FIR
at $t\sim 1$ Gyr. We see this point again later in Section \ref{subsec:Z_nearby}.

In Fig.\ \ref{fig:sed}, we also examine the dependence on $\eta_\mathrm{dense}$.
We observe that the FIR peak intensity
is not sensitive to $\eta_\mathrm{dense}$. This is because the FIR luminosity is determined by
the total dust mass, which is not sensitive to $\eta_\mathrm{dense}$.
In contrast, the intensity of the MIR emission, which originates from
small grains including PAHs, is sensitive to $\eta_\mathrm{dense}$.
This is because the small-grain abundance and the aromatic fraction are high in the ISM dominated by
the diffuse medium (Section \ref{subsec:size}).
{We note that the SED slope at long wavelengths is constant under a given dust species.
Thus, the change of SED shape is prominent at $\lambda\lesssim 200~\micron$, which
further justifies our usage of
the \textit{Spitzer} band coverage for the comparison with nearby galaxies.}

We also see the silicate feature at 9.7 $\micron$ in the early epochs (especially
for $\eta_\mathrm{dense}=0.5$ and 0.9) in Fig.\ \ref{fig:sed}. This is because the silicate fraction
is higher in the
early epoch. For $\eta_\mathrm{dense}=0.9$, PAH emission is weak because the aromatic fraction is low.
Weaker PAH features make the 9.7 $\micron$ emission relatively prominent. Therefore, dense and young
conditions are favourable for a prominent silicate feature.

Finally, as mentioned in Section \ref{subsec:obs}, we compare the resulting SEDs with the Milky
Way data. In the case with $\eta_\mathrm{dense}=0.5$ (Fig.~\ref{fig:sed}b), we observe that the
long-wavelength data points are reproduced with ages of a few Gyr, while the data at short
wavelengths where PAH emission is prominent is consistent with the results at $\sim 10$ Gyr.
This means that the SEDs reproducing the Milky Way FIR emission tends to underpredict the PAH emission.
This could be resolved if we assume a lower value of $\eta_\mathrm{dense}$. Indeed, as shown
in Fig.\ \ref{fig:sed}a, the intensities at short wavelengths are no longer underproduced
at $t\sim{}$ a few Gyr. However, the emission around $\lambda =20$--70 $\micron$ is overpredicted,
since the small-grain abundance is significantly enhanced for small $\eta_\mathrm{dense}$.
For $\eta_\mathrm{dense}=0.9$, the intensities at $\lambda\lesssim 30~\micron$ are underproduced
because of the suppressed small-grain abundance.
Any values of $\eta_\mathrm{dense}=0$--1 have difficulty in reproducing the SEDs
at $\lambda =20$--70 $\micron$ and at $\lambda\lesssim 10~\micron$ simultaneously.
However, we emphasize that the Milky Way SED data are successfully covered by our predictions with
$\eta_\mathrm{dense}=0.1$--0.5 at $t\sim{}$ a few--10 Gyr.
The difficulty in reproducing the data with a single
parameter set should be resolved in a future work; for example,
a superposition of different grain size distributions in different
ISM phases would be necessary to reproduce the Milky Way SED at all wavelengths at once.
Alternatively, the grain size distribution is different from what we predict; indeed, DL07
proposed that the size distribution of PAHs has log-normal components.
Such a log-normal grain size distribution is difficult to reproduce in our
model.
As shown later, the difficulty in simultaneously reproducing the multi-band data
arises also for the nearby galaxy sample, so that
we revisit this problem again in Section \ref{sec:discussion}.

\subsection{Dependence on the ISRF intensity}\label{subsec:U}

\begin{figure}
\includegraphics[width=8cm]{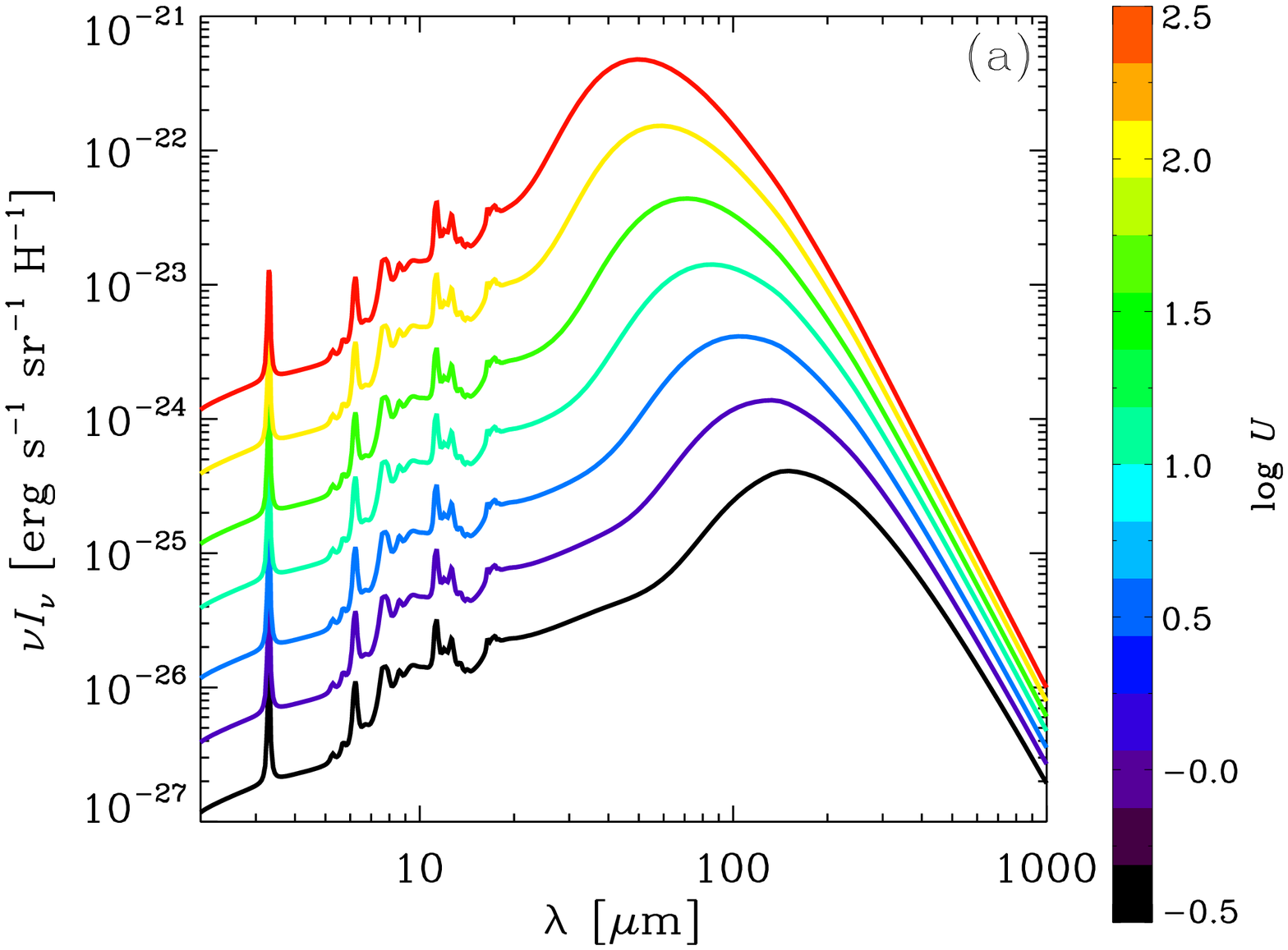}\\
\includegraphics[width=8cm]{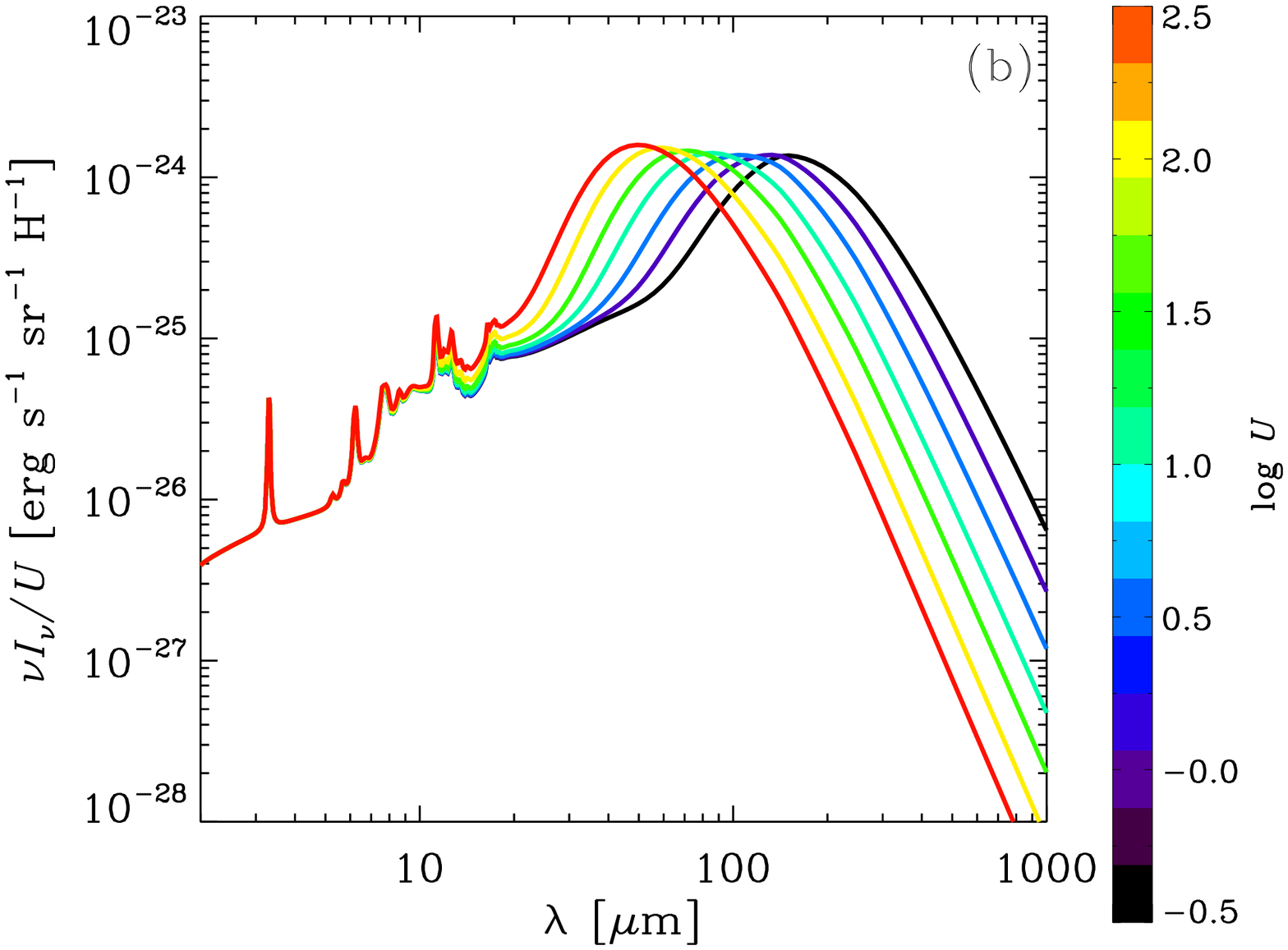}
\caption{SEDs at $t=10$ Gyr for various values of $U$.
We adopt $\eta_\mathrm{dense}=0.5$, $\tau_\mathrm{SF}=5$ Gyr, and the DL07 grain
compositions.
Panels (a) and (b) show the intensity itself and the intensity divided by $U$, respectively.
We show the results for $U=0.3$, 1, 3, 10, 30, 100, and 300
in colours shown in the colour bar.
\label{fig:sed_U}}
\end{figure}

In this paper, we simply treat the ISRF intensity as a free parameter.
Here we examine the effect of $U$ on the
SED. Since the effect of $U$ is qualitatively
similar for any grain size distributions and many previous studies have already investigated
the effect of $U$ on the IR SEDs (e.g.\ \citealt{Draine:1985aa,Li:2001aa}; DL07),
we focus on the SEDs at $t=10$ Gyr with
$\eta_\mathrm{dense}=0.5$ and $\tau_\mathrm{SF}=5$ Gyr as shown in Fig.\ \ref{fig:sed_U}.
As expected, the emission at all wavelengths becomes stronger as $U$ increases. The SED peak in the FIR
shifts to shorter wavelengths as the ISRF becomes stronger because the equilibrium dust temperature becomes
higher. To cancel out the natural increase of the overall dust emission with $U$,
we also show the intensity divided by $U$.
As noted by previous studies (e.g.\ DL07),
we find that the MIR SED, where the PAH emission is
prominent, is simply scaled with $U$ (i.e.\ the intensity divided by $U$ is constant).
This is explained by the nature of stochastic heating in which
the hitting frequency of photons on small grains
is proportional to $U$.
On the other hand, the FIR SED peak, which is determined by the
equilibrium temperature, shifts
to shorter wavelengths as $U$ increases.
The maximum value of $\nu I_\nu$, which roughly reflects the total
IR luminosity, still scales with $U$; this means that the
total IR luminosity (= total reprocessed luminosity from dust) also
scales with $U$ ($\propto$ dust heating by stellar radiation).
This is a natural consequence of radiative equilibrium.
Therefore, the MIR SED and the total IR luminosity scale with $U$, while the SED peak
shifts to shorter wavelengths for higher $U$. This peak shift is the main reason that the
intensity ratios at various wavelengths changes with $U$. This fact is important in interpreting the
intensity ratios shown below.

\subsection{Evolution with metallicity}\label{subsec:Z_nearby}

Since we are interested in galaxy evolution, it is crucial to adopt an indicator of
the evolutionary stages.
Since the galaxy age is difficult to determine, we often use the metallicity as an indicator of galaxy
evolution \citep[e.g.][]{Tinsley:1980aa}. Here we focus on some quantities characterizing the SED shape
as a function of metallicity and compare them with
the data described in Section \ref{subsec:obs}.
In our model, the metallicity $Z$ is calculated by the chemical evolution model described in
Section \ref{subsec:chem}.
We focus on the case with $\tau_\mathrm{SF}=5$~Gyr and vary $\eta_\mathrm{dense}$ and $U$.
We discuss the dependence on $\tau_\mathrm{SF}$ in Section \ref{subsec:tau}.
In our model, the star formation time-scale $\tau_\mathrm{SF}$ also gives the time-scale
on which the metallicity reaches solar ($\sim 1$~Z$_{\sun}$).
{We show the evolution up to $t=10$ Gyr.}


\subsubsection{FIR intensity ratio}

\begin{figure}
\includegraphics[width=8cm]{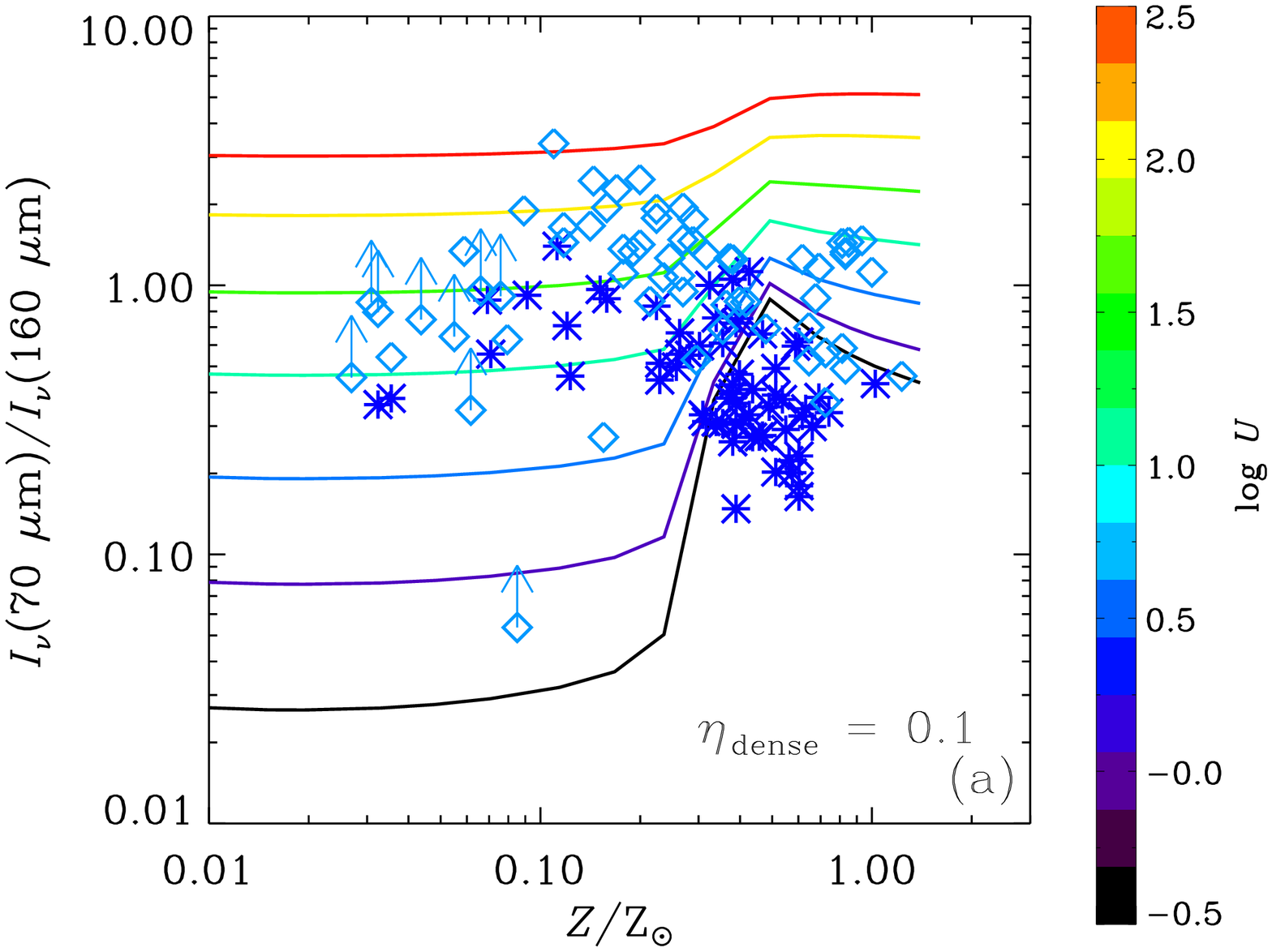}\\
\includegraphics[width=8cm]{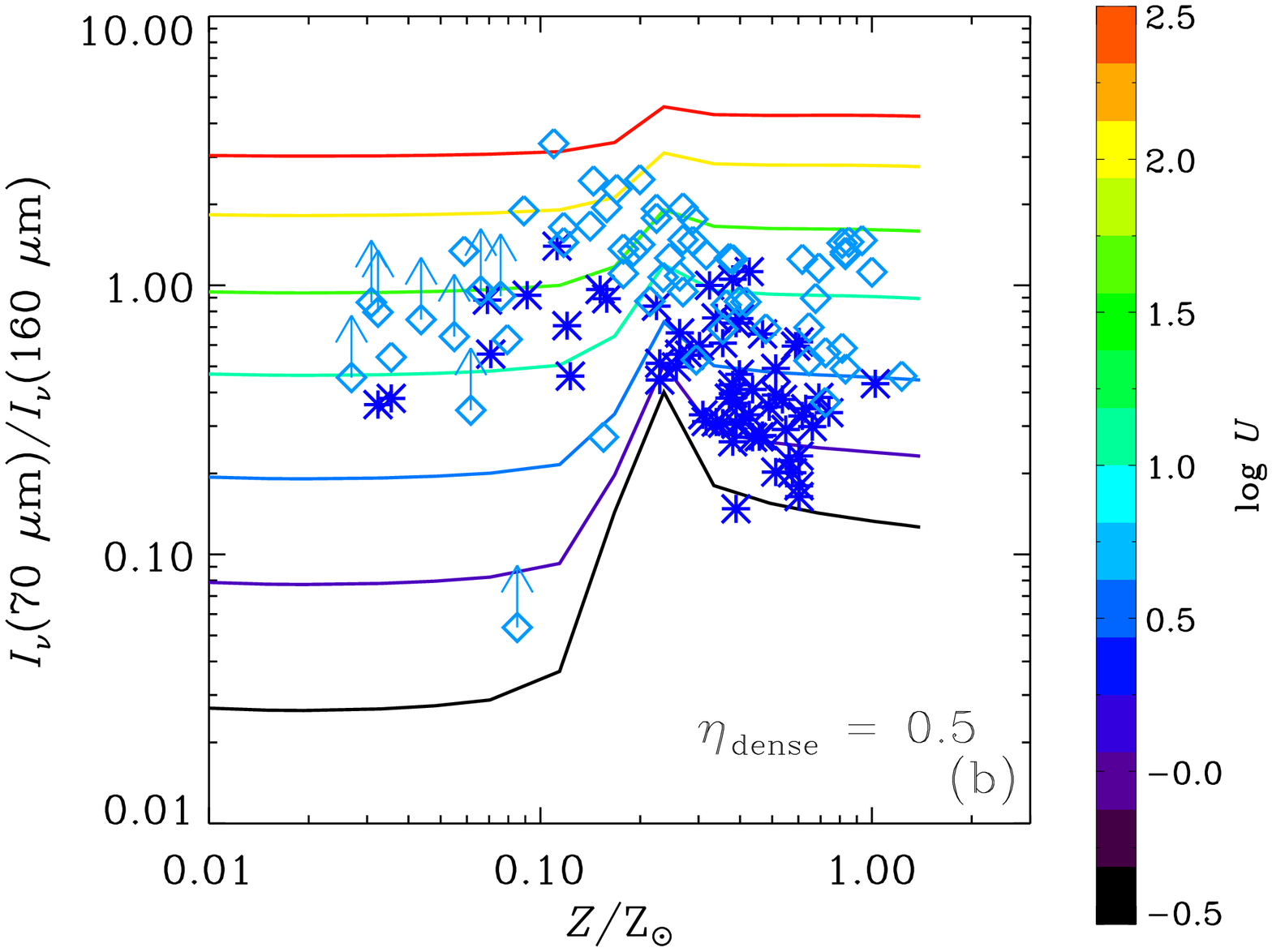}\\
\includegraphics[width=8cm]{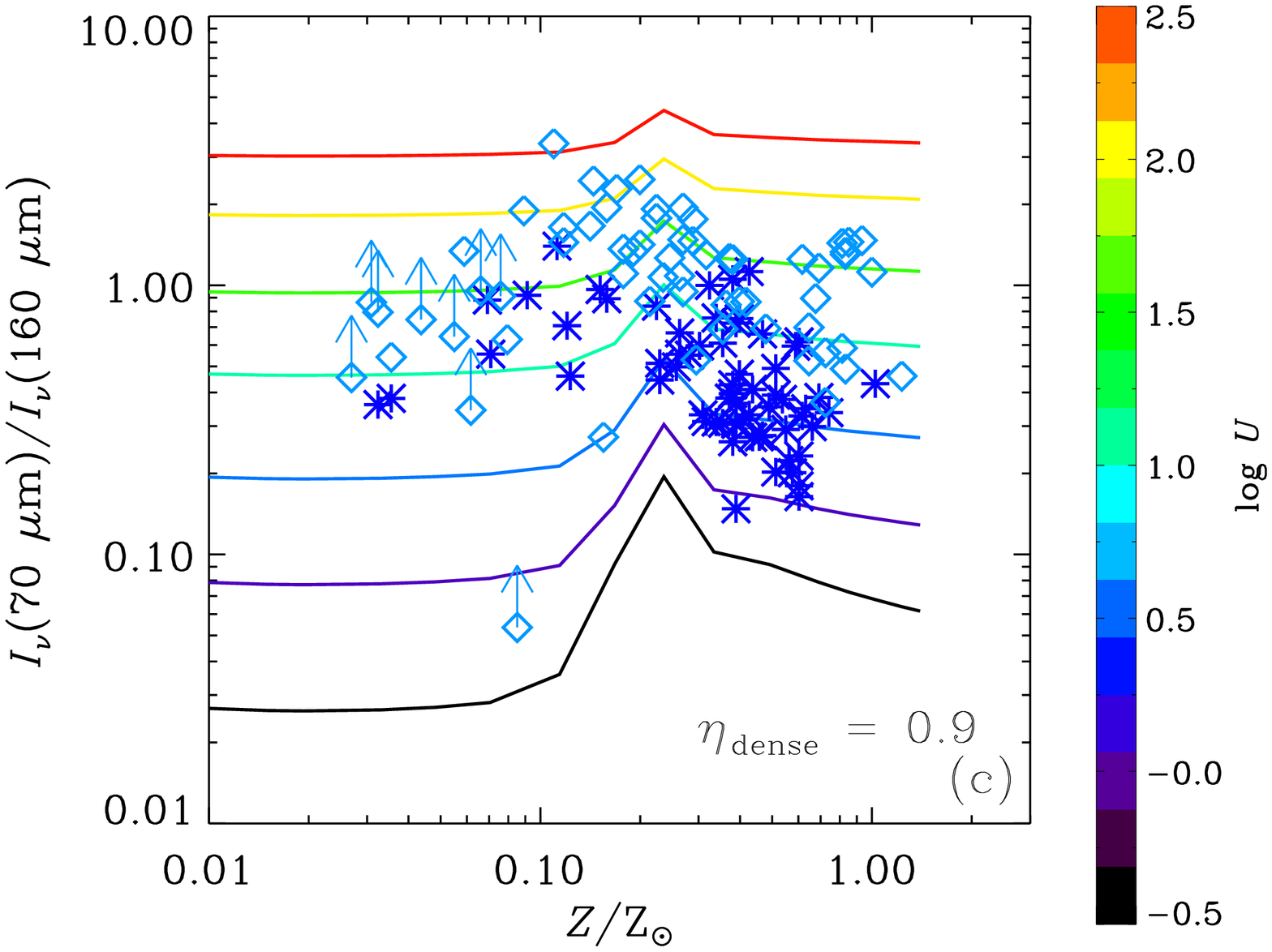}
\caption{70--160 $\micron$ ratio as a function of metallicity
for $\eta_\mathrm{dense}=0.1$, 0.5, and 0.9
in Panels (a), (b), and (c), respectively
(with $\tau_\mathrm{SF}=5$ Gyr and the DL07 grain compositions).
The colours show the same values of $U$ as in Fig.\ \ref{fig:sed_U}.
The observational data are taken from \citet{Dale:2017aa} and
\citet{Engelbracht:2008aa} (asterisks and diamonds, respectively).
The arrows show lower limits.
\label{fig:70_160}}
\end{figure}

Although the ISRF intensity is treated as a free parameter in this paper, we can obtain
a reasonable range for it by inspecting the bands which reflect the equilibrium dust temperature.
Among the \textit{Spitzer} bands, the 70 and 160~$\micron$ bands,
which are the nearest to the SED peak,
reflect the equilibrium dust temperature the most (Section~\ref{subsec:U}).
In Fig.~\ref{fig:70_160}, we plot $I_\nu (70~\micron )/I_\nu (160~\micron )$
(referred to as the 70--160~$\micron$ ratio) as a function of metallicity
for $\eta_\mathrm{dense}=0.1$, 0.5, and 0.9 and for $U=0.3$--300.
First, we focus on the case with $\eta_\mathrm{dense}=0.5$ (Fig.\ \ref{fig:70_160}b).
As expected, the 70--160 $\micron$ ratio becomes larger for higher $U$.
However, there is also a change in the intensity ratio along each line, where $U$ is constant.
This is because the 70 $\micron$ intensity
is still affected by stochastically heated small grains \citep{Bernard:2008aa}.
{For example, in the case of M33, the contribution from such small grains
to the 70 $\micron$ intensity is $\sim 30$ per cent \citep{Relano:2018aa}.}
The relatively suppressed
70--160 $\micron$ intensity
ratio at low metallicity is due to the lack of small grains (recall that stellar sources, which
dominate the dust production at low metallicity, produce large grains).
The peak around $Z\simeq 0.2$ Z$_{\sun}$ corresponds to the epoch
when the small-grain abundance is drastically enhanced by accretion.
At high metallicity, there is a slight decreasing trend for the 70--160 $\micron$ ratio
because small grains are
efficiently converted to large grains by coagulation.

We explain the observational data of nearby galaxies with $U\sim 1$--30 for
metal-rich objects, confirming previous studies \citep[e.g.][]{Draine:2007ab}.
Metal-poor objects tends to have higher $U$ ($\sim 10$--100). This confirms that dust in
low-metallicity ($Z\sim 0.2~\mathrm{Z}_{\sun}$) star-forming galaxies tend to be irradiated
by intense radiation
\citep[e.g.][]{Engelbracht:2008aa,Hirashita:2009ac,Hunt:2010aa}.
It is interesting to point out that the 70--160 $\micron$ ratios of the observational sample
seem to have a peak around $Z\sim 0.1$--0.2 Z$_{\sun}$ as already noted by \citet{Engelbracht:2008aa}.
In our model, this peak is at least partly explained by the increase of small grains by accretion.

The evolution of the 70--160 $\micron$ ratio also depends on $\eta_\mathrm{dense}$.
As we observe in Fig.\ \ref{fig:70_160}, the difference between
the cases with $\eta_\mathrm{dense}=0.5$ and 0.9 is small,
because as shown in HM20, the evolution of
grain size distribution is similar except for the
overabundance of large grains at $a\sim 1~\micron$
for $\eta_\mathrm{dense}=0.9$.
In contrast, the 70--160 $\micron$ ratio is significantly higher
for $\eta_\mathrm{dense}=0.1$ than for the other cases at
high metallicity since the grain size distribution is dominated by small grains.

\subsubsection{MIR}

Now we examine the MIR wavelength range, where the emission is dominated by small grains.
Among the \textit{Spitzer} bands, the 8 and 24 $\micron$ bands can be used to
examine the contribution from small grains.
The 8 $\micron$ emission is sensitive to the PAH emission, while the 24~$\micron$ band is
more affected by the entire small-grain
population  (Section \ref{subsec:obs}). We use the TIR intensity for the normalization.

\begin{figure}
\includegraphics[width=0.45\textwidth]{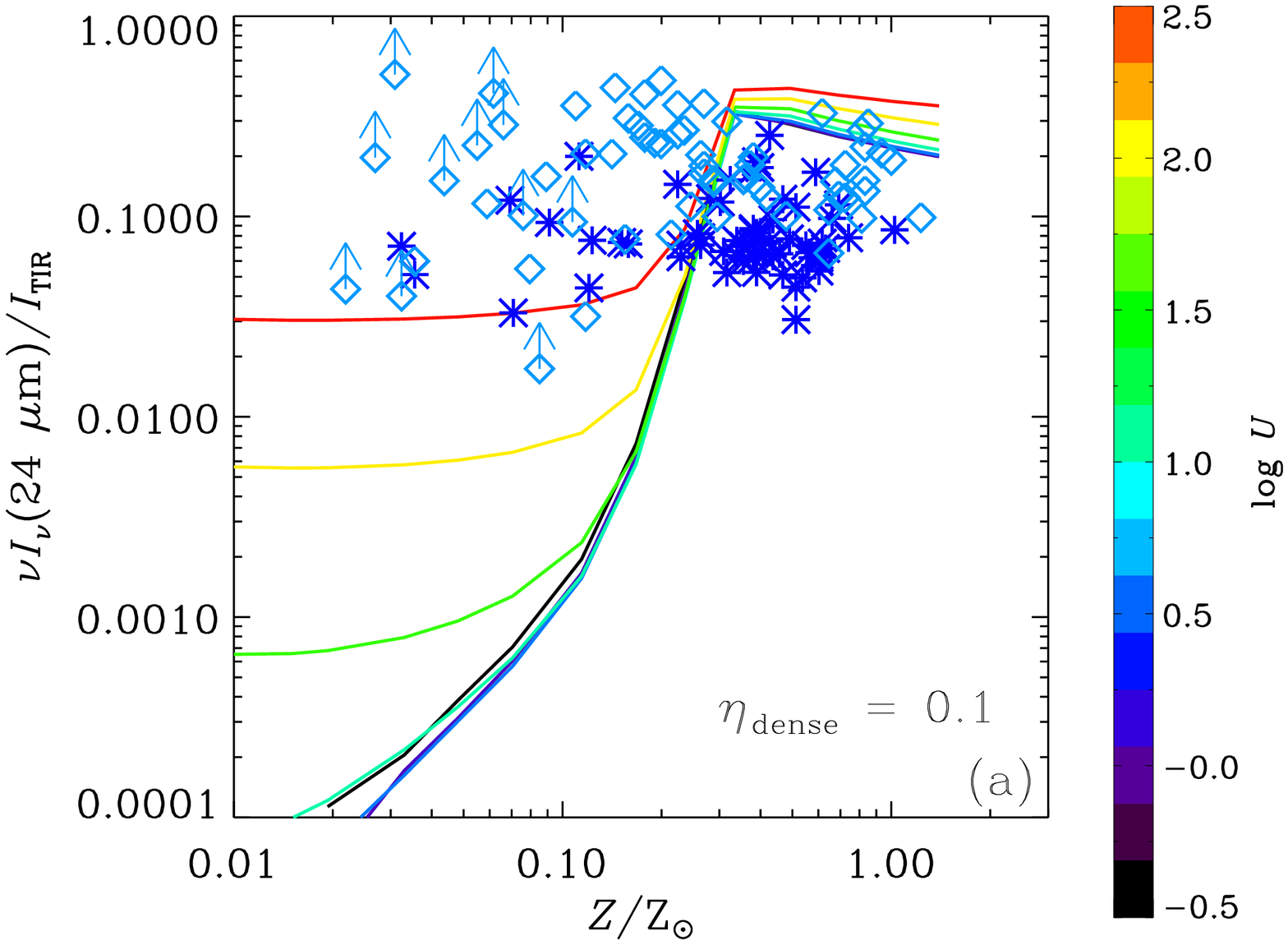}\\
\includegraphics[width=0.45\textwidth]{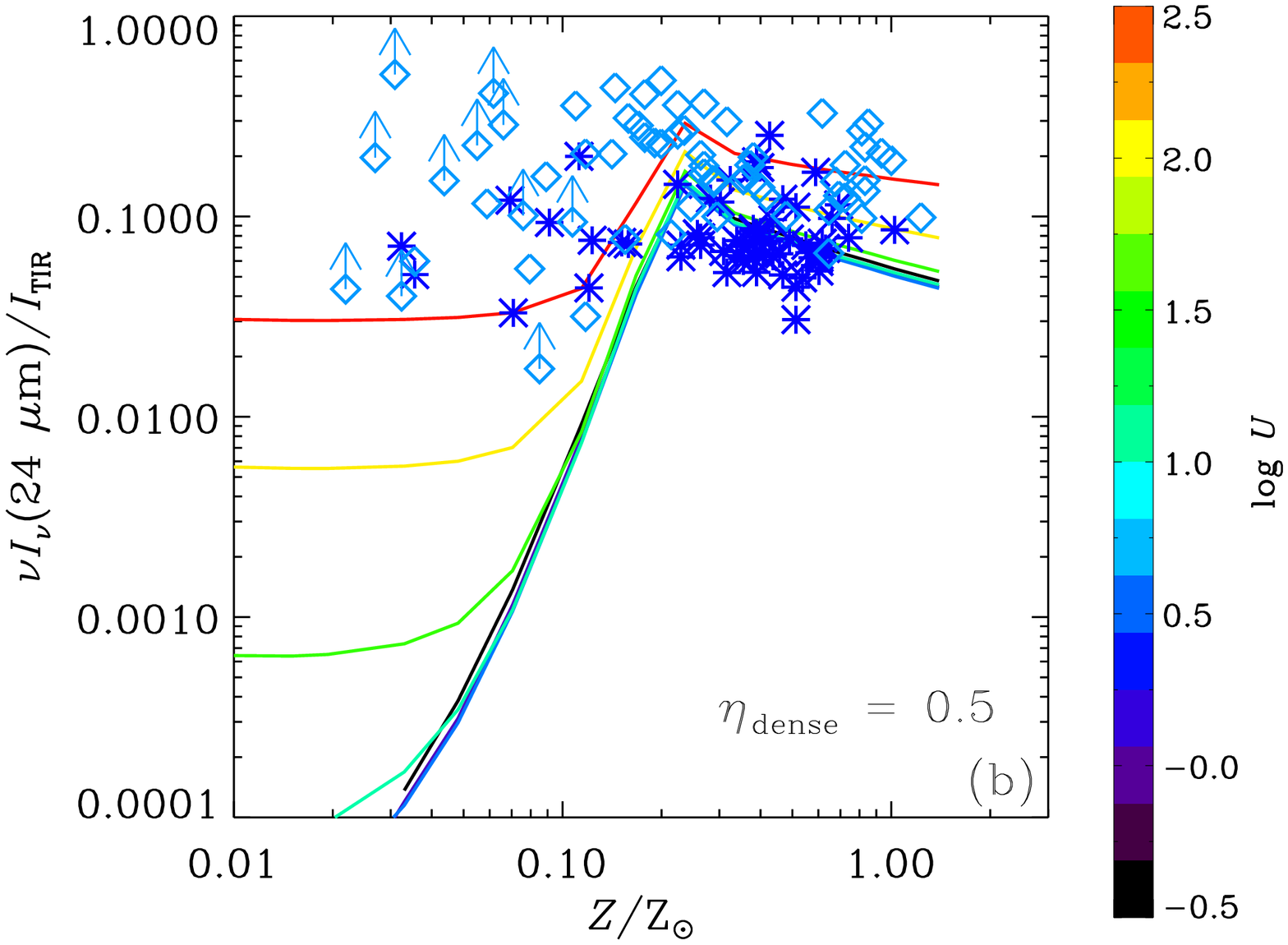}\\
\includegraphics[width=0.45\textwidth]{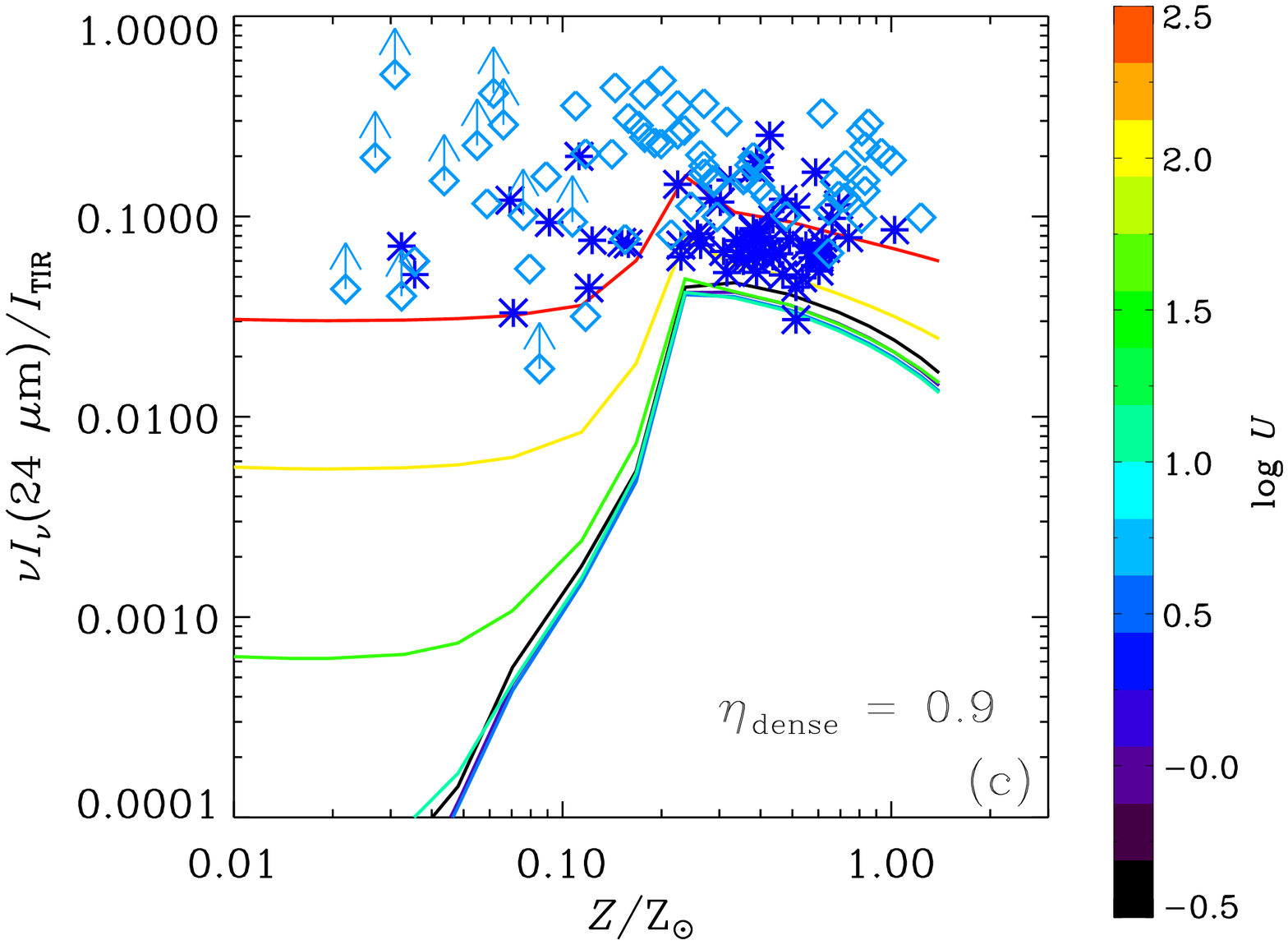}
\caption{Same as Fig.\ \ref{fig:70_160} but for the 24 $\micron$--TIR ratio.
The lines are almost completely overlapping for low values of $U$.
\label{fig:24_TIR}}
\end{figure}

First, we show in Fig.~\ref{fig:24_TIR} the evolution of $\nu I_\nu (24~\micron)/I_\mathrm{TIR}$,
which is referred to as the 24 $\micron$--TIR ratio.
The 24 $\micron$--TIR ratio increases with metallicity at $Z\lesssim 0.2$ Z$_{\sun}$,
which reflects the small-grain production by shattering and accretion. At high metallicity,
it decreases because coagulation converts small grains to large grains.
We observe that the results are not sensitive to $U$ for $U\lesssim 100$ at high metallicity,
since both 24 $\micron$ and TIR intensities scale with $U$ almost linearly (Section \ref{subsec:U}).
If $U$ is as strong as 300, large grains, which are in radiative equilibrium,
affect the 24 $\micron$ emission. This radiative-equilibrium component breaks the
scaling of $I_\nu$ with $U$ at a fixed wavelength as shown in Section \ref{subsec:U}.
At low metallicity, because the abundance of small grains is small, large grains could
easily dominate the 24 $\micron$ (i.e.\  the 24 $\micron$ intensity is affected by the
radiative equilibrium component, which does not simply scale with $U$).
Thus, for the same reason as above, the 24 $\micron$--TIR
ratio depends on the ISRF intensity at low metallicity.

From Fig.\ \ref{fig:24_TIR}, we also find that
the results are somewhat sensitive to $\eta_\mathrm{dense}$ especially at high metallicity.
As expected from the SEDs shown in Fig.\ \ref{fig:sed},
a larger $\eta_\mathrm{dense}$ tends to suppress the MIR emission more because of
more efficient coagulation and less efficient shattering.
At high metallicity ($Z\gtrsim 0.2$~Z$_{\sun}$), the 24 $\micron$--TIR ratio is high for
a smaller value of $\eta_\mathrm{dense}$.
It is worth pointing out that the medium case ($\eta_\mathrm{dense}=0.5$) explains the
observational data at high metallicity {the most among the three values of
$\eta_\mathrm{dense}$.
The decreasing trend at high metallicity predicted by our model is not inconsistent with
the data.
This decreasing trend is caused by coagulation, which converts small grains (responsible for the
24 $\micron$ emission) to large grains.}
We also observe that the scatter in the observational data at high metallicity is also explained
by the variation of $\eta_\mathrm{dense}$.
{Since the scatter is large, a large range of $\eta_\mathrm{dense}$
(such as 0.1 for the upper part of the data points, and 0.9 for the lower part) is consistent with the
data at high metallicity. The ISRF intensity causes the dispersion only if $U\gtrsim 300$ for some galaxies,
but such a high ISRF is not supported by the 70--160 $\micron$ ratio for high-metallicity
galaxies (Fig.\ \ref{fig:70_160}).}
At low metallicity, since the dust abundance is dominated by stellar dust production,
the result is not sensitive to $\eta_\mathrm{dense}$.
The metallicity level at which the 24 $\micron$--TIR ratio peaks is not sensitive
to $\eta_\mathrm{dense}$; this is because the metallicity, not $\eta_\mathrm{dense}$,
is the main quantity that determines the efficiency of dust growth by accretion
\citep{Inoue:2011aa}.
Thus, the data points with high 24 $\micron$--TIR ratios at low metallicity is
difficult to explain by the change of $\eta_\mathrm{dense}$.
{In Section \ref{subsec:tau}, we investigate a possibility of resolving this
discrepancy by longer $\tau_\mathrm{SF}$.}

\begin{figure}
\includegraphics[width=8cm]{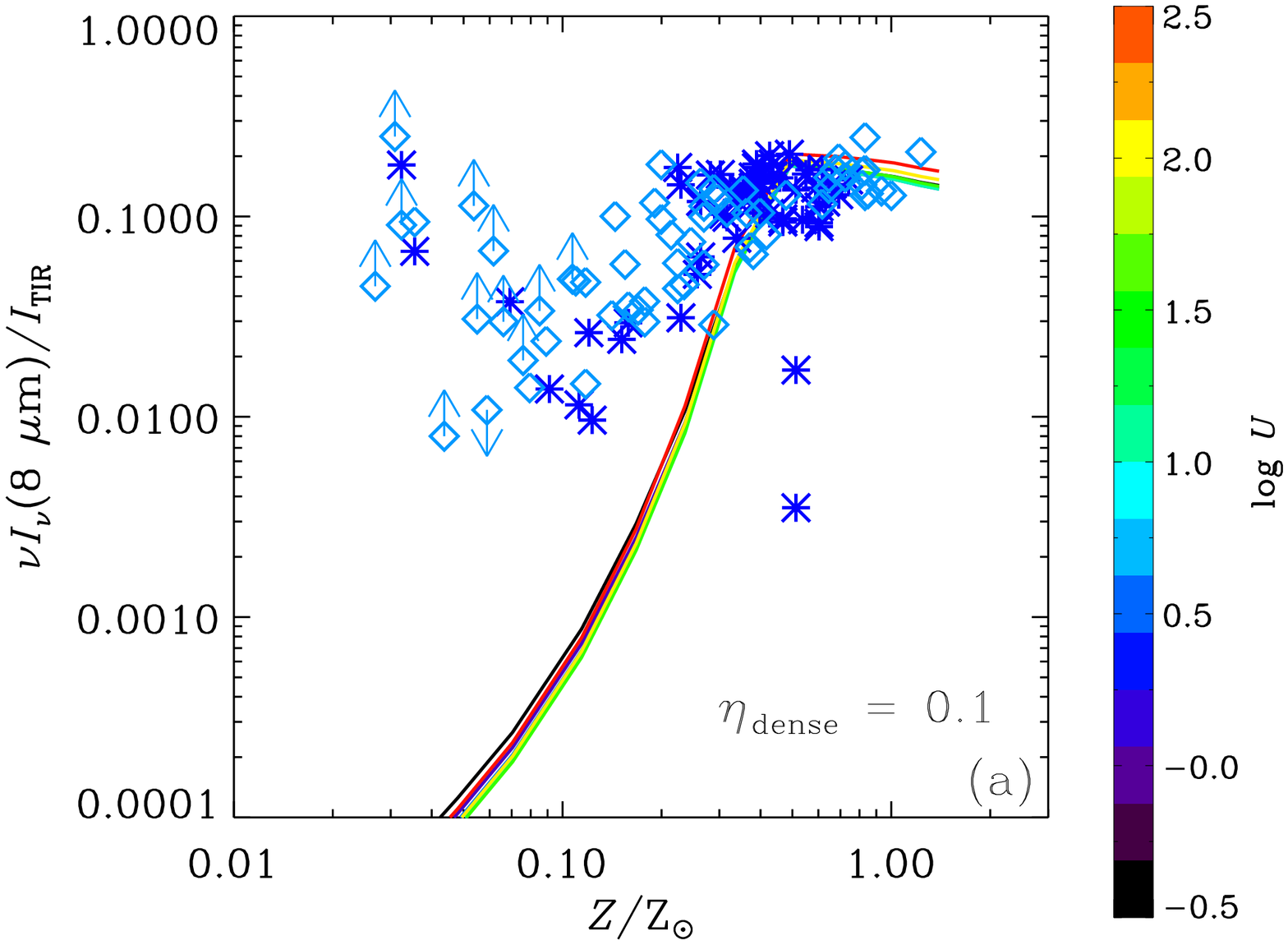}\\
\includegraphics[width=8cm]{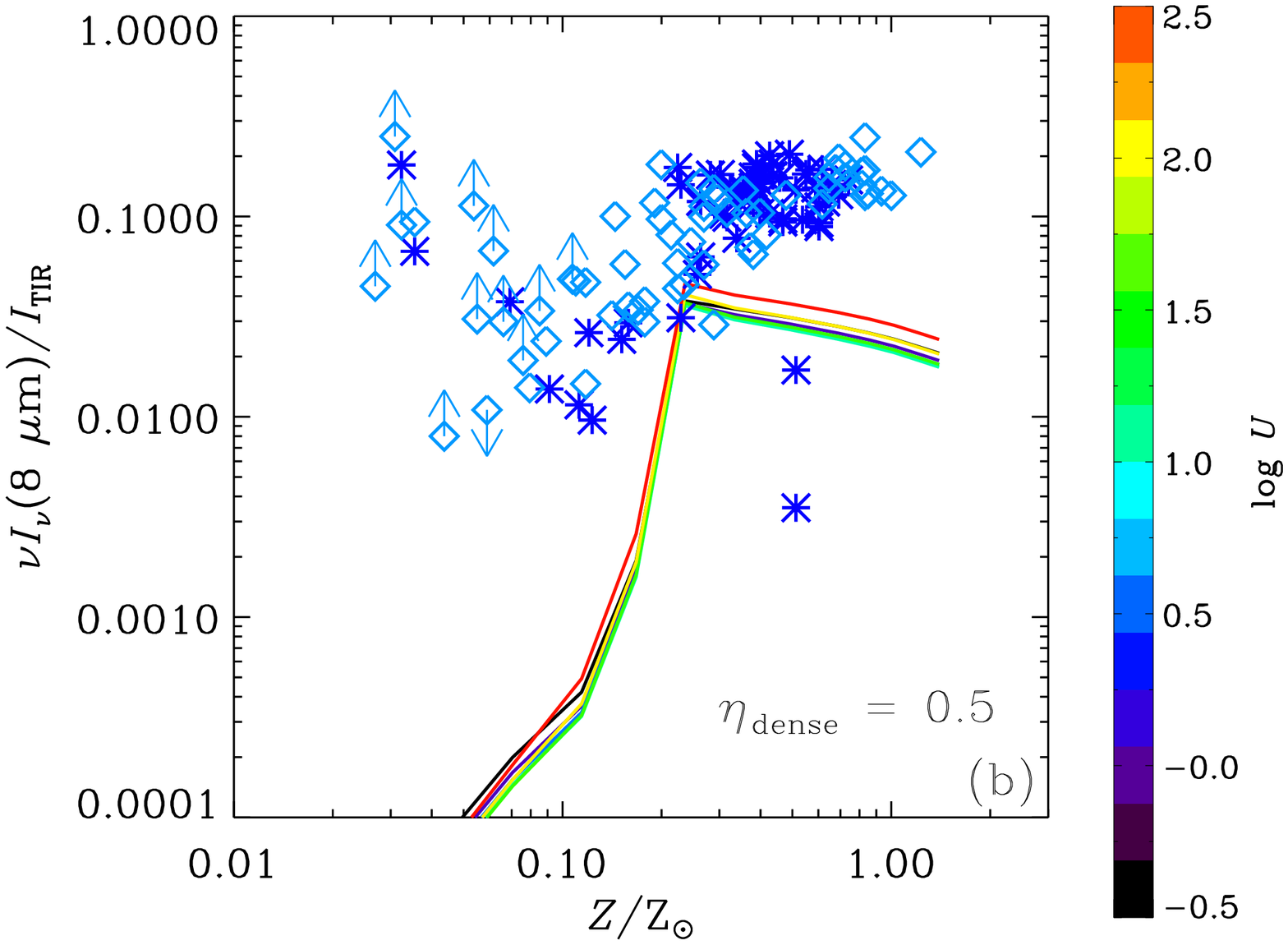}\\
\includegraphics[width=8cm]{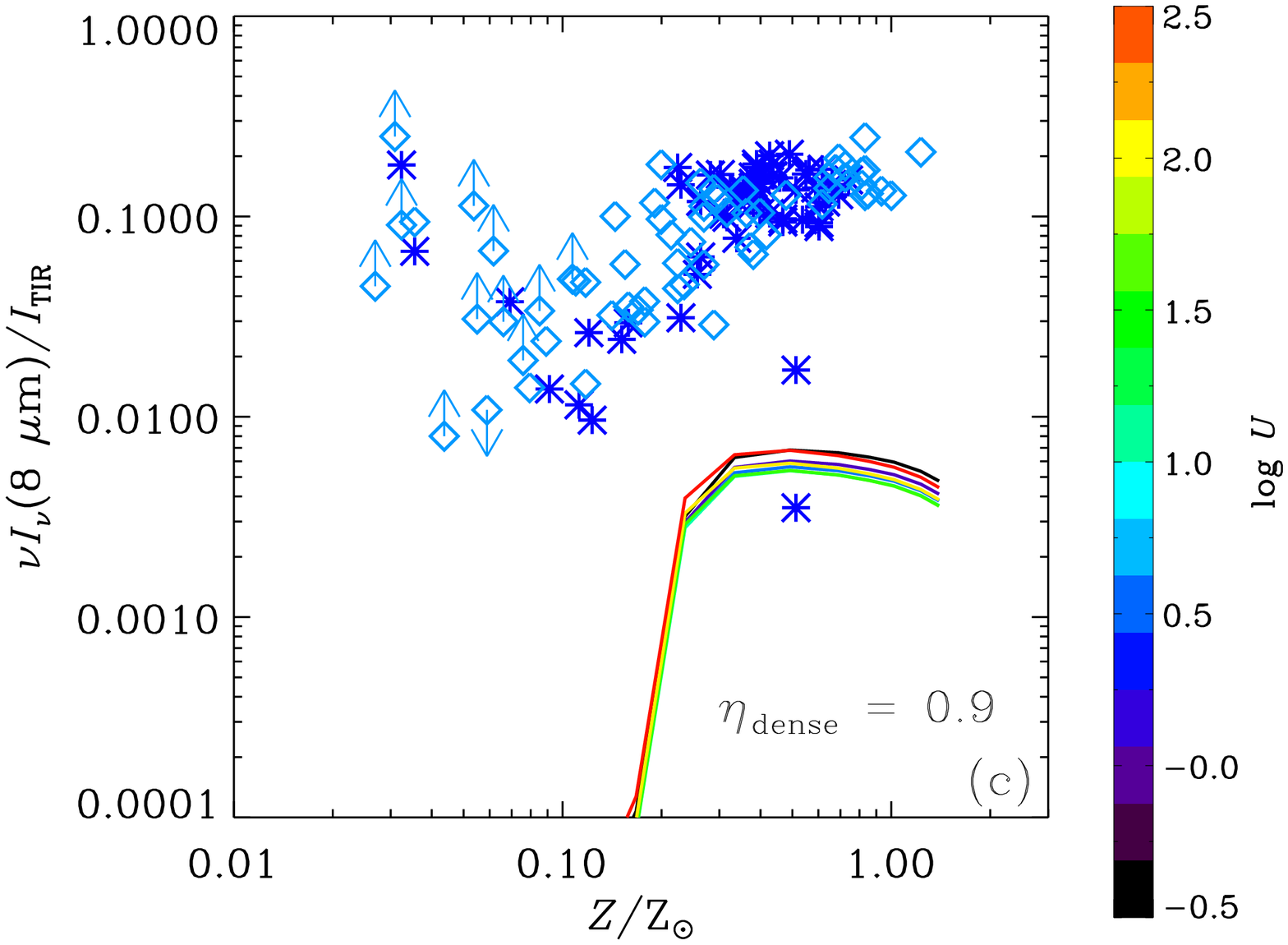}
\caption{Same as Fig.\ \ref{fig:70_160} but for the 8 $\micron$--TIR ratio.
The upward and downward arrows show lower and upper limits, respectively.
\label{fig:8_TIR}}
\end{figure}

Next, we examine $\nu I_\nu (8~\micron )/I_\mathrm{TIR}$
(referred to as the $8~\micron$--TIR ratio) in Fig.~\ref{fig:8_TIR}.
As mentioned above, the $8~\micron$ band is chosen as a band sensitive to the PAH emission.
We observe that the 8~$\micron$--TIR ratio behaves in a similar way as the
24~$\micron$--TIR ratio, since both of these two MIR bands reflect the abundance of
small grains relative to that of large grains.
Thus, the above interpretations for the 24 $\micron$--TIR ratio are mostly valid for the
8 $\micron$--TIR ratio. There are some differences, though.
Since the 8 $\micron$ band is far from the SED peak even for the case of $U=300$, it is not
affected by the radiative equilibrium component (i.e.\ large grains).
Thus, the 8 $\micron$--TIR ratio is not sensitive to $U$ in the entire metallicity range.
The observation data at high metallicity are located near the theoretical prediction
with $\eta_\mathrm{dense}=0.1$, favouring the ISM dominated by the diffuse medium.
Recall that the curves with $\eta_\mathrm{dense}=0.5$ {are located in the middle of}
the data {at high metallicity} for the 24 $\micron$--TIR ratio.
Thus, the favoured $\eta_\mathrm{dense}$ value is different between the 8 and 24 $\micron$ bands.
Probably, the 8 $\micron$ and 24 $\micron$ emissions are dominated by different ISM phases; thus,
spatially resolved treatments for dust evolution
would be necessary (see also Section \ref{subsec:enhance_PAH}).
All the curves underpredict the
observed 8 $\micron$--TIR ratio at low metallicity ($Z\lesssim 0.2$ Z$_{\sun}$).
The fact that the 8 $\micron$--TIR ratio is predicted to be insensitive to $U$ means that
the scatter in the observational data is caused not by the ISRF intensity,
but by other properties such as the dense gas fraction (see above) and the star formation time-scale
(see Section \ref{subsec:tau}).
{The underprediction of the 8 $\micron$--TIR ratio at low metallicity is not resolved even if
we change $\eta_\mathrm{dense}$ with metallicity, since all values of $\eta_\mathrm{dense}$
underproduce the 8 $\micron$--TIR ratio. This is resolved at least
partially by a longer star formation time-scale, but some data points with high
8 $\micron$--TIR ratios would need additional mechanism of PAH formation that is not included in
this paper. This issue is further discussed in Sections \ref{subsec:resolved_PAH} and \ref{subsec:enhance_PAH}.}

\begin{figure}
\includegraphics[width=8cm]{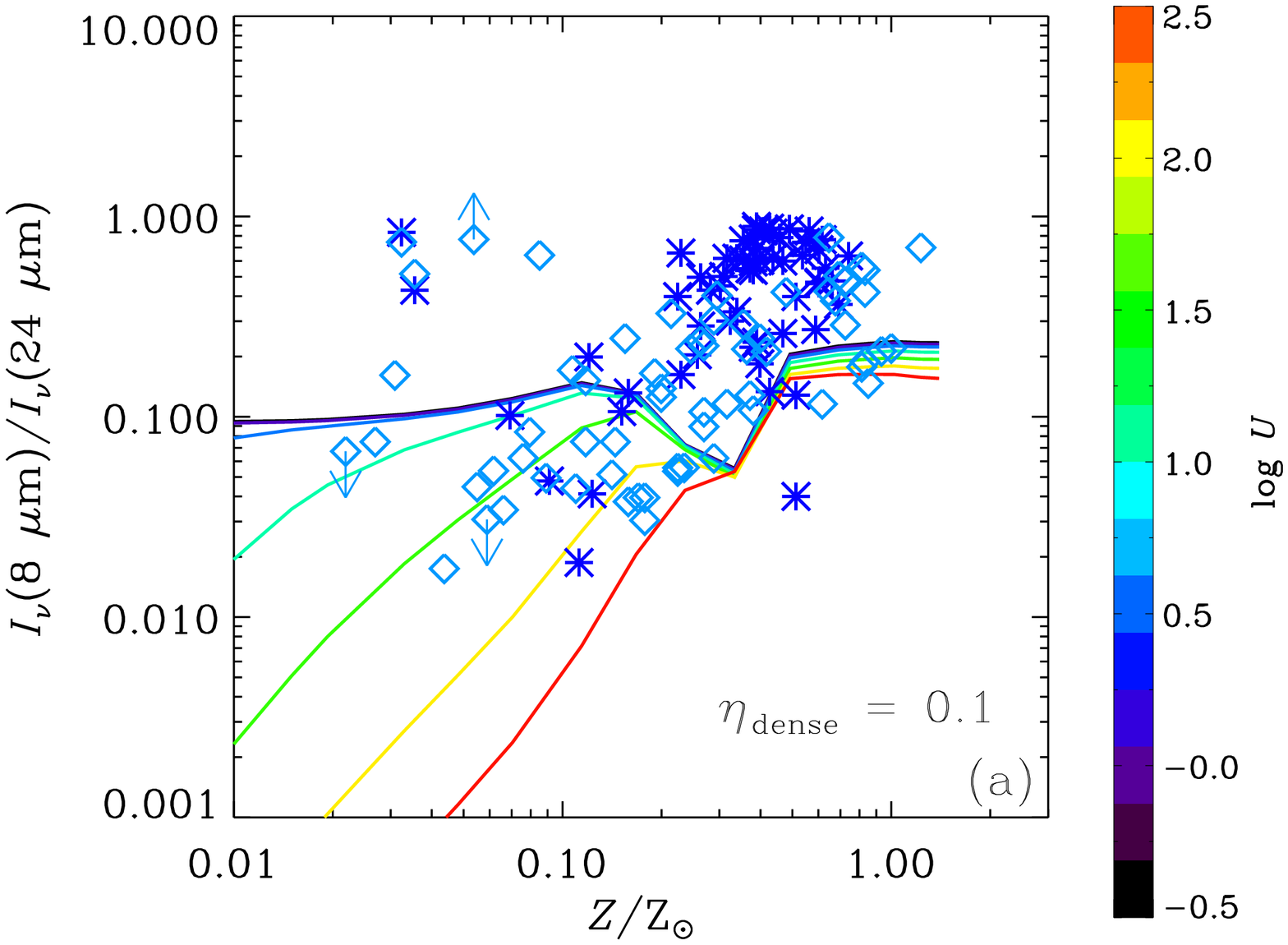}\\
\includegraphics[width=8cm]{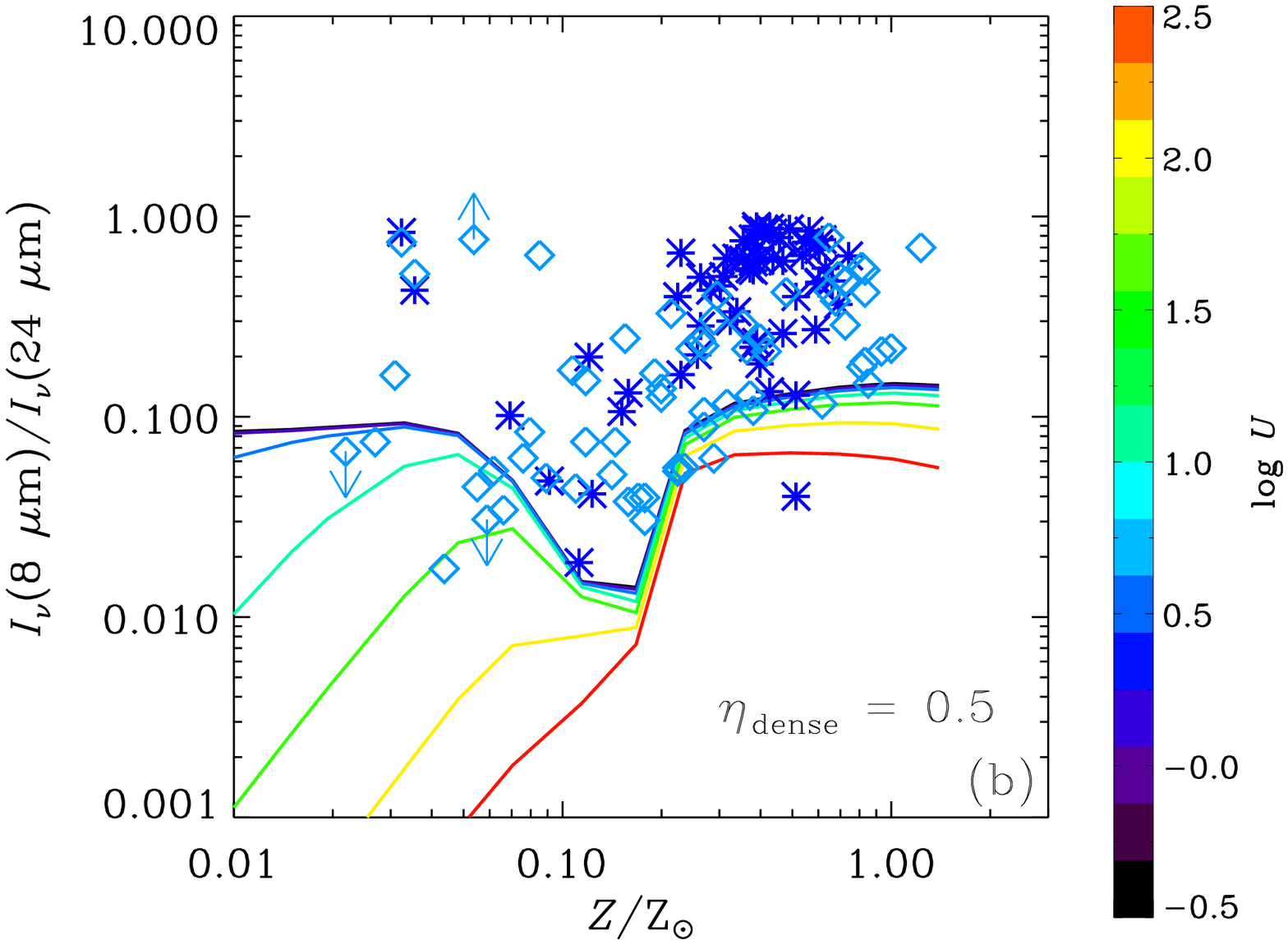}\\
\includegraphics[width=8cm]{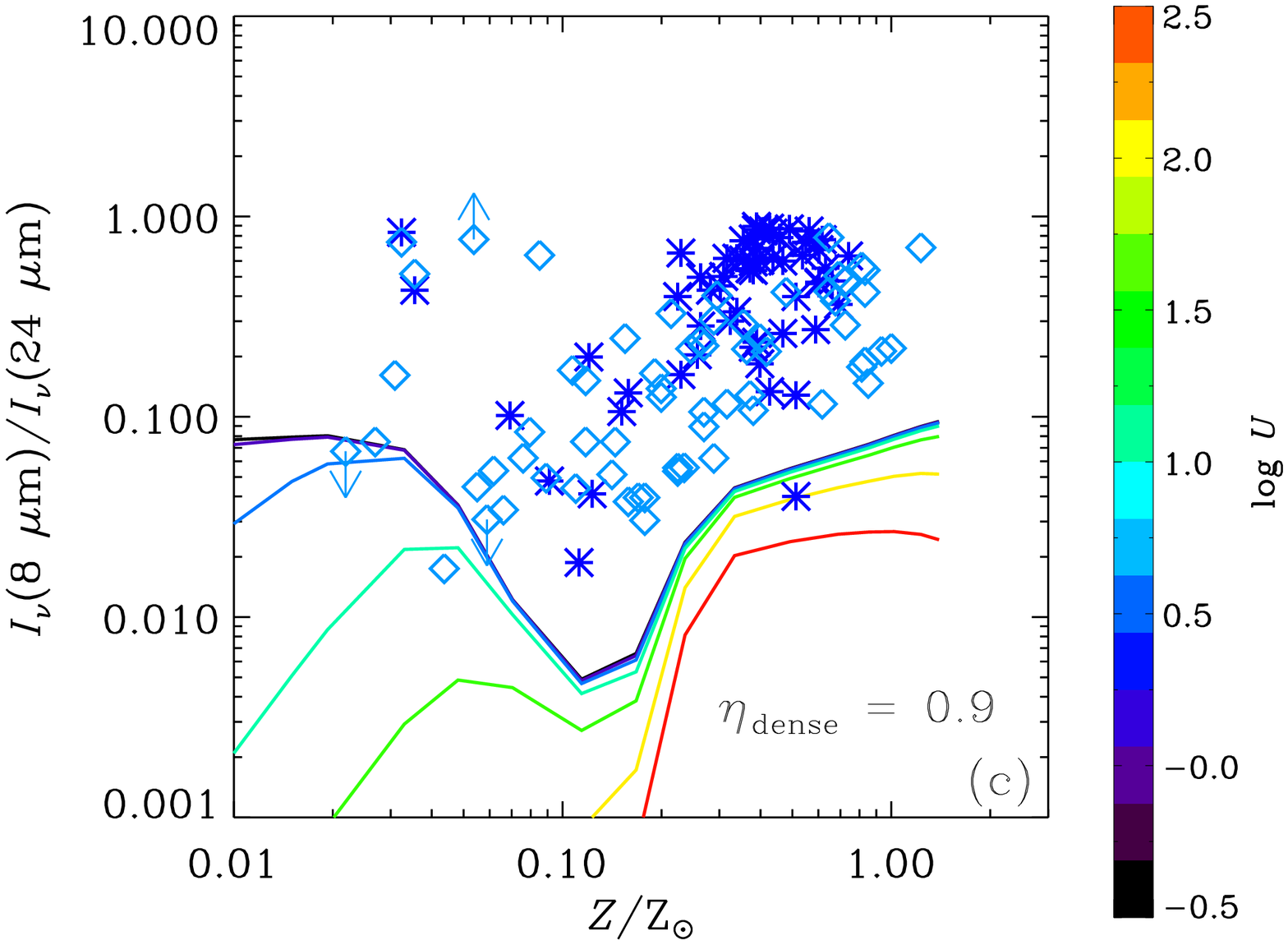}
\caption{Same as Fig.\ \ref{fig:8_TIR} but for the 8--24 $\micron$ ratio.
\label{fig:8_24}}
\end{figure}

It is also interesting to investigate $I_\nu (8~\micron )/I_\nu (24~\micron )$ (referred to as
the 8--24~$\micron$ ratio), since this ratio reflects the PAH fraction in the small-grain population.
We show the 8--24 $\micron$ ratio in Fig.\ \ref{fig:8_24}.
At low metallicity, the ratio depends strongly on $U$ because
the 24~$\micron$ intensity is raised by the contribution from large grains for high $U$.
For all values of $U$, the 8--24 $\micron$ ratio increases at $Z\gtrsim 0.2$ Z$_{\sun}$,
and is not sensitive to the variation of $U$ in that metallicity range.
The 8--24 $\micron$ ratio tends to be suppressed for
higher $\eta_\mathrm{dense}$ because the aromatic fraction is lower.
All cases tend to underpredict the observational data at high metallicity.
This confirms the above `discrepancy' between the favoured values of
$\eta_\mathrm{dense}$ in the 8 $\micron$--TIR and 24 $\micron$--TIR ratios.

\subsection{Different star formation time-scales}\label{subsec:tau}

As shown by HM20, the star formation time-scale $\tau_\mathrm{SF}$ also affects the
evolution of dust and PAH abundances with metallicity.
The characteristic metallicity at which the abundances of PAHs and small grains increase rapidly shifts
towards lower metallicities as $\tau_\mathrm{SF}$ becomes longer.
Thus, the evolution of the MIR intensities is expected to be strongly affected
by the change of $\tau_\mathrm{SF}$.
We focus on the intensity ratios that involve the MIR bands
(the 24~$\micron$--TIR, 8 $\micron$--TIR, and 8--24 $\micron$ ratios) to investigate the effect
of $\tau_\mathrm{SF}$.

\begin{figure}
\includegraphics[width=8cm]{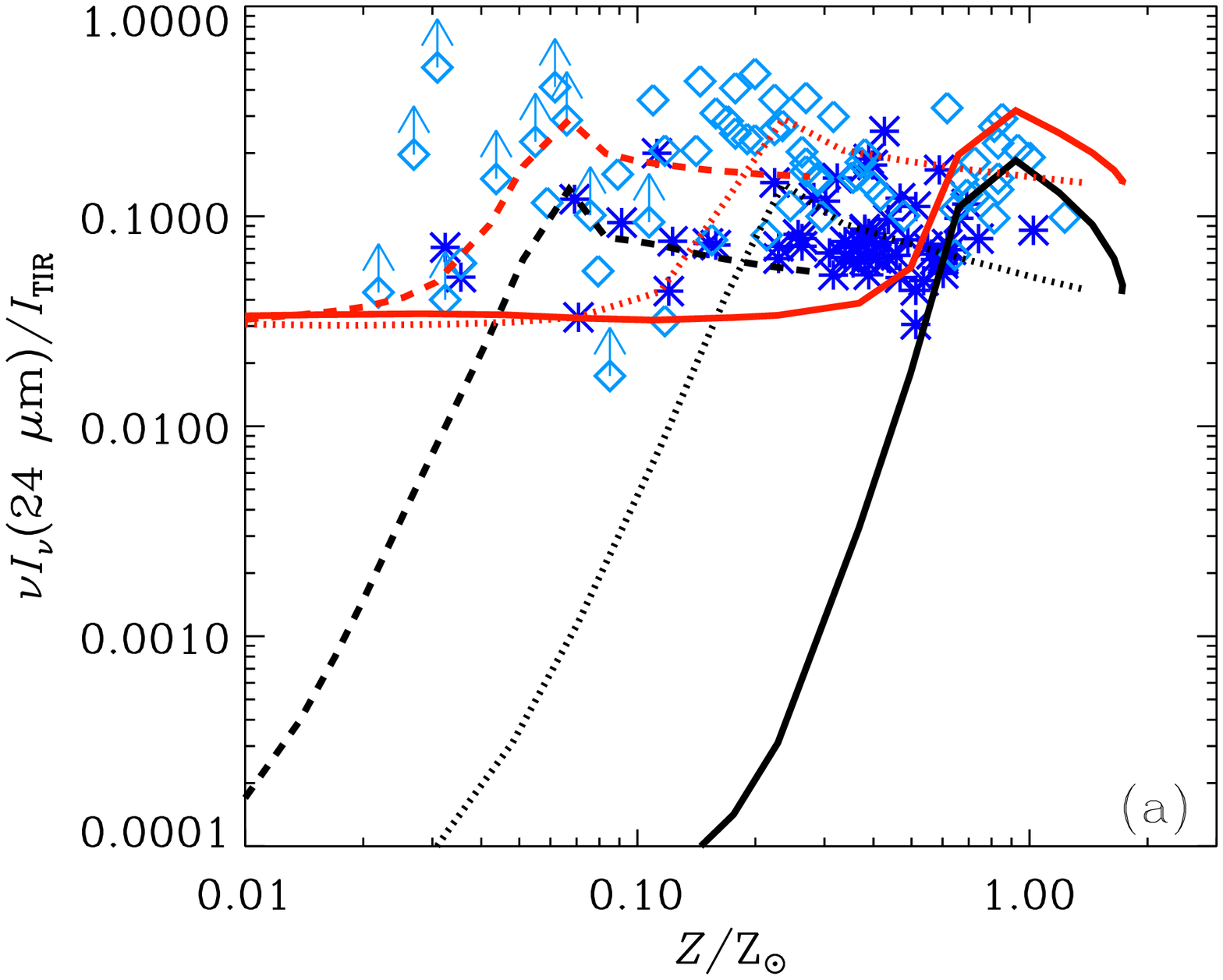}\\
\includegraphics[width=8cm]{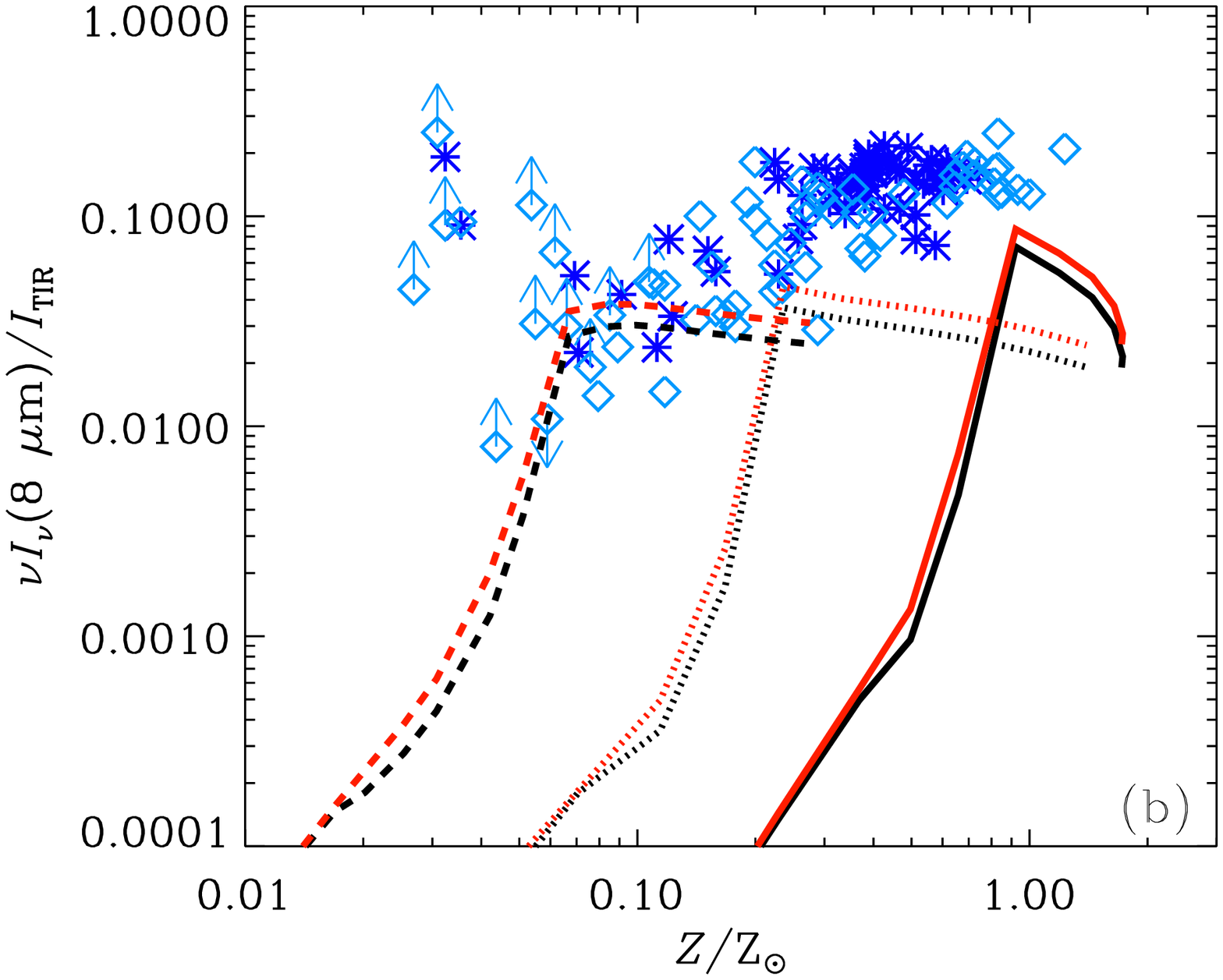}\\
\includegraphics[width=8cm]{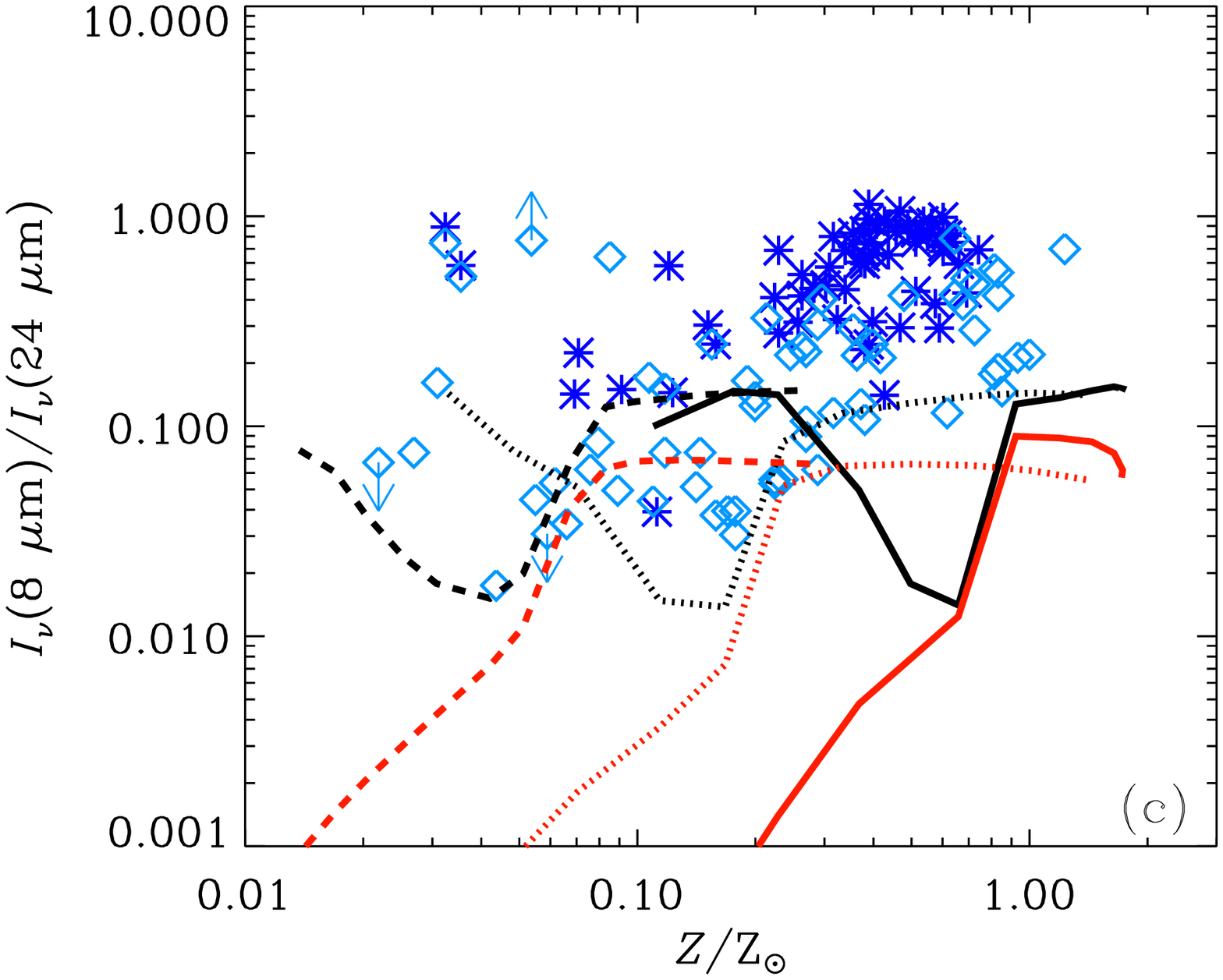}
\caption{24 $\micron$--TIR, 8 $\micron$--TIR, and 8--24 $\micron$ ratios
(Panels a, b, and c, respectively) for $\tau_\mathrm{SF} = 0.5$, 5, and 50 Gyr
(solid, dotted, and dashed lines, respectively)
for $U=1$ and 300 (black and red lines, respectively).
\label{fig:clr_tau}}
\end{figure}

We show the results for $\tau_\mathrm{SF}=0.5$, 5, and 50~Gyr in Fig.~\ref{fig:clr_tau}.
{For $\tau_\mathrm{SF}=0.5$ Gyr, we stop the calculation at $t=3$ Gyr because
most of the gas is consumed. For the other cases, we calculate the evolution up to $t=10$ Gyr.}
We only present the cases with $U=1$ and 300.
We observe that the evolutionary tracks shift to lower metallicities as
$\tau_\mathrm{SF}$ becomes longer. This is because the
metallicity level at which the dust abundance is rapidly raised by accretion scales as
$\propto\tau_\mathrm{SF}^{-1/2}$ \citep{Asano:2013ab}:
If the chemical enrichment occurs quickly, the system is more enriched with metals
before the interstellar processing of dust becomes efficient.
A long $\tau_\mathrm{SF}$ could help to explain some of the data points
at low metallicity by realizing a high value of the 24 $\micron$--TIR and
8 $\micron$--TIR ratios at low metallicity. However, since the difference in $\tau_\mathrm{SF}$
only causes a horizontal shift in the diagrams,
the high 8--24 $\micron$ ratios in the observational data cannot be explained
by the change of $\tau_\mathrm{SF}$. In summary, the different values of $\tau_\mathrm{SF}$
serve to produce horizontal scatters (in the direction of metallicity)
on the intensity ratio--metallicity diagrams.

\section{Discussion}\label{sec:discussion}

\subsection{Dust properties from J13}\label{subsec:J13}

In the above, we used the DL07 model for the optical properties (or grain compositions).
The J13 model, which we also adopted in HM20, is another well investigated set of
dust properties (Section \ref{subsec:SED}).
The J13 model is characterized by a detailed modelling of
carbon dust properties,
especially covering a full variation for the band gap energy of a-C(:H).
Our motivation of adopting J13 in this paper is to investigate the robustness of the prediction
against the change of dust optical properties.
Note that we still use the same grain size distributions as used above.

\begin{figure}
\includegraphics[width=8cm]{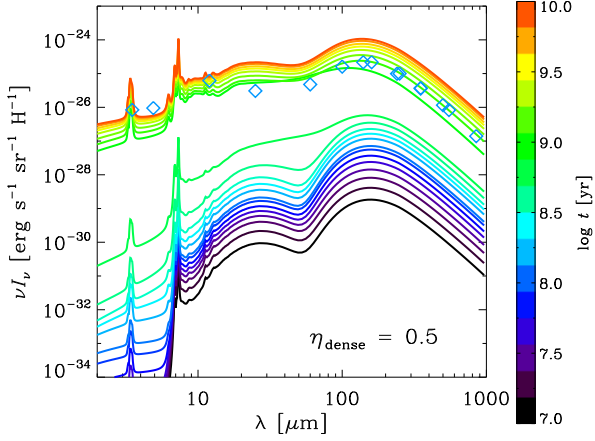}
\caption{Same as Fig.\ \ref{fig:sed}(b) but with the J13 model for the dust optical properties.
\label{fig:sed_j13}}
\end{figure}

In Fig.\ \ref{fig:sed_j13}, we show the evolution of SED for the J13 model.
We fix $\eta_\mathrm{cold}=0.5$, $\tau_\mathrm{SF}=5$~Gyr, and $U=1$.
This figure is to be compared with Fig.\ \ref{fig:sed}(b), where we adopted the DL07 grain compositions.
We find that the detailed shape of the MIR spectral features is significantly different
between the two models. However, we should note that
there is a lot of room for the adjustment of detailed dust
properties such as the surface coating, etc.\ \citep{Jones:2017aa}. We emphasize again that the
purpose of adopting the J13 model here is to examine the sensitivity of the results against the
change of dust optical properties, not to judge which model is better.
There is another large difference in the MIR. Because
the non-aromatic grains have extremely low
infrared emissivity \citep{Jones:2012aa,Ysard:2018aa}, their equilibrium temperatures are high.
This creates a MIR bump at $\lambda\sim 20$--30 $\micron$ even without small grains.
Consequently, the J13 model predicts stronger MIR emission than the DL07 model, especially
in the early epoch, when the small-grain abundance is low.
The SEDs at $\lambda\gtrsim 70~\micron$ do not differ significantly between the two models.
Thus, the FIR SED is robust against the change of the dust properties.

We also plot the observational data for the Milky Way SED in Fig.\ \ref{fig:sed_j13}.
The data points in the FIR are reproduced with ages of a few Gyr, while the data
at short wavelengths where PAH emission is prominent is consistent with
the results at $t\sim 10$ Gyr.
This means that the PAH emission is underestimated if we choose an age
($t\sim\text{a few Gyr}$) that explains the observed FIR SED.
This underestimate also occurred in the DL07 model.
Indeed, except for the detailed MIR feature shapes,
the overall SED shapes from the MIR to the FIR at $t\gtrsim\text{a few Gyr}$ are
similar between the J13 and DL07 models (compare Fig.\ \ref{fig:sed_j13} with
Fig.\ \ref{fig:sed}b).
Thus, as far as the broad SED shape at later ages is concerned,
the prediction is robust against the change of the dust optical properties.

\begin{figure}
\includegraphics[width=8cm]{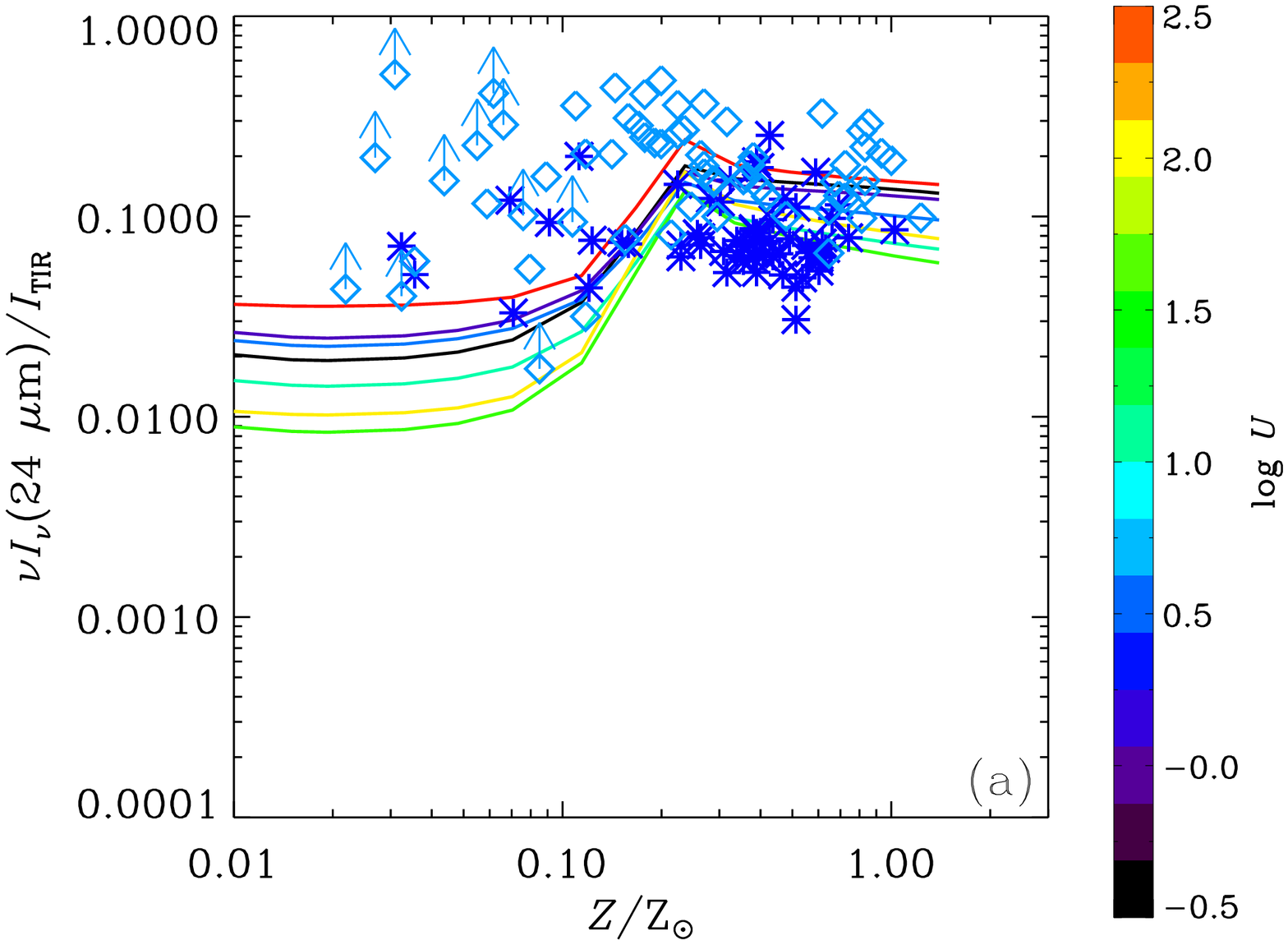}\\
\includegraphics[width=8cm]{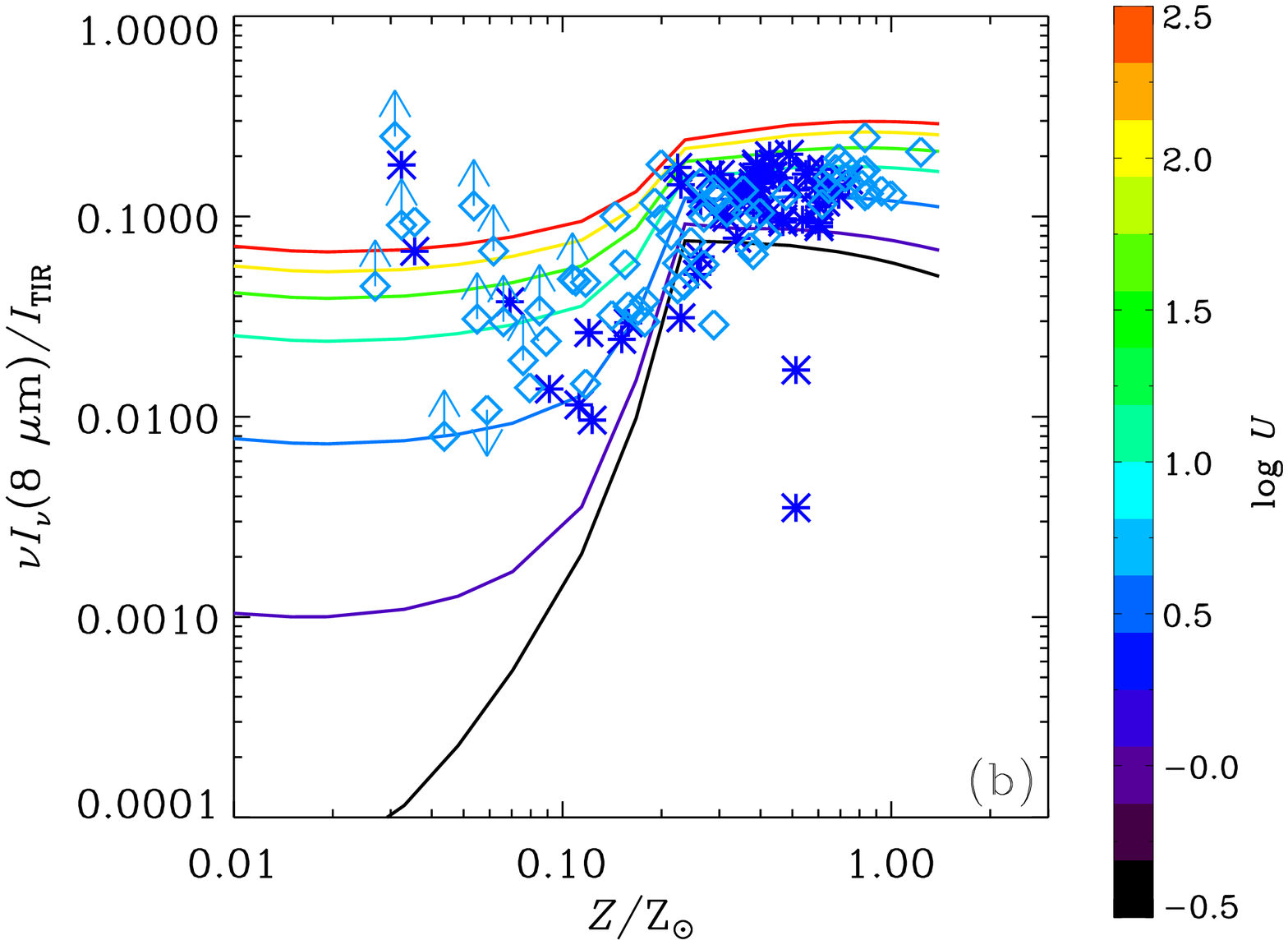}\\
\includegraphics[width=8cm]{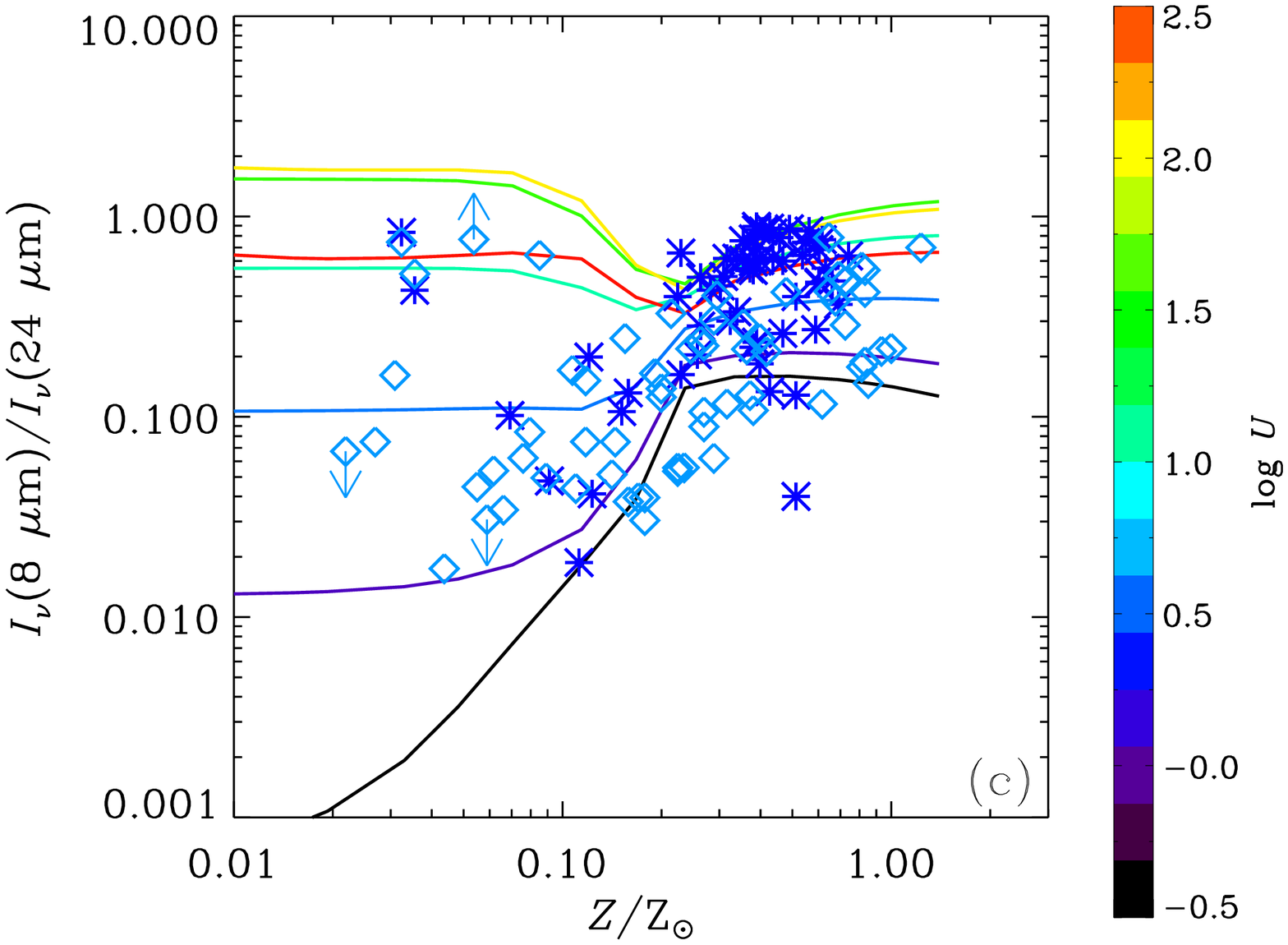}
\caption{Intensity ratios predicted with the J13 model for the
dust optical properties (with
$\eta_\mathrm{dense}=0.5$ and $\tau_\mathrm{SF}=5$ Gyr). Panels (a), (b), and (c)
show the 24 $\micron$--TIR, 8 $\micron$--TIR, and 8--24 $\micron$ ratio, respectively,
for $U=0.3$, 1, 3, 10, 30, 100, and 300, whose colour coding is shown by the colour bar.
\label{fig:clr_j13}}
\end{figure}

Since the difference between the two models appear in the MIR, we show
the evolution of the 24~$\micron$--TIR, 8~$\micron$--TIR and 8--24~$\micron$ ratios
in Fig.\ \ref{fig:clr_j13}. Because the emission at $\lambda\sim 20$--30 $\micron$
is enhanced, the 24~$\micron$--TIR ratio is high at low metallicity
compared with the DL07 model (Fig.~\ref{fig:clr_j13}a vs.\ Fig.~\ref{fig:24_TIR}b).
The difference in the 24 $\micron$--TIR ratio between the J13 and DL07 models is small
at high metallicity because small grains become the major contributor to
the 24 $\micron$ emission in both cases.
The 8~$\micron$--TIR ratios in the J13 model are overall higher
than those in the DL07 model (Fig.\ \ref{fig:clr_j13}b vs.\ Fig.~\ref{fig:8_TIR}b)
because of the contribution from the tail of the above 20--30 $\micron$ excess.
Since this tail is sensitive to the dust temperature
determined by the ISRF intensity ($U$), the 8 $\micron$--TIR ratio changes
for different values of $U$ in the J13 model; in the DL07 model, in contrast,
the ratio is almost insensitive to $U$.
Because of the larger variation of the 8 $\micron$--TIR ratio in the J13 model,
the scatter of the observational data is better explained at both high and low
metallicities. For the same reason, the observed 8--24 $\micron$ ratios are covered
better in the J13 model than in the DL07 model (Fig.\ \ref{fig:8_24}b).

As mentioned above, we do not intend to conclude which of the DL07 and J13 models is better.
We conservatively state that the J13 model covers the observational range better than the
DL07 model with the grain size distributions predicted by our model.
We note that even in the DL07 model,
the prediction with small $\eta_\mathrm{dese}$
could also reproduce the observational data (Fig.\ \ref{fig:8_TIR}a).

\subsection{Comparison with $z\sim 2$ galaxies}

The 8~$\micron$ band used to trace PAH emission at $z\sim 0$ is redshifted to another
\textit{Spitzer} (24~$\micron$) band at $z\sim 2$. Here we compare our model with
the observational data taken from \citet{Shivaei:2017aa} (at $z=1.4$--2.6)
because they also show the relation with
the metallicity estimated from rest-frame optical metal lines.
They stacked the IR data of the sample in a few metallicity bins
(40--90 galaxies with a median redshift of 2.1--2.3 in each metallicity bin),
and derived the rest-frame 7.7~$\micron$ luminosity and the total FIR luminosity as well as
other observational quantities for the sample.
They used the \citet{Chary:2001aa} SED template
to derive the monochromatic luminosity at 7.7 $\micron$. Thus, we calculate
the ratio between the 7.7 $\micron$ intensity multiplied by the frequency corresponding to
$\lambda =7.7~\micron$ and the TIR intensity. This ratio is denoted as
$\nu L_\nu (7.7~\micron )/L_\mathrm{TIR}$, and referred to as the 7.7 $\micron$--TIR ratio.

\begin{figure}
\includegraphics[width=8cm]{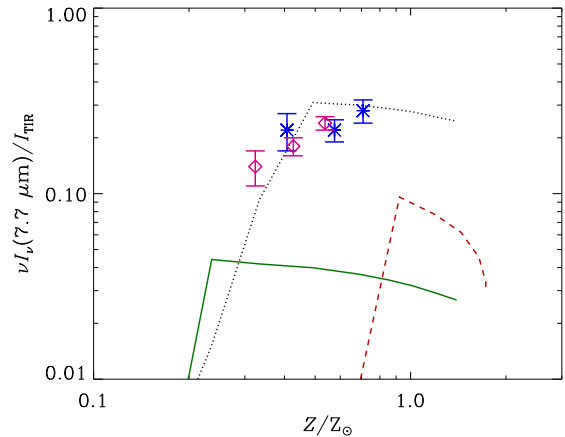}
\caption{7.7 $\micron$--TIR ratio as a function of metallicity to be compared with the
$z\sim 2$ sample. The solid, dotted, and dashed lines show the results for
$(\eta_\mathrm{dense},\,\tau_\mathrm{SF})=(0.5,\, 5~\mathrm{Gyr})$,
(0.1, 5 Gyr), and (0.5, 0.5 Gyr), respectively.
We adopt the stacked $z\sim 2$ data from \citet{Shivaei:2017aa}, who used
two different metallicity determination methods: the asterisks and diamonds show
the data with
different metallicity estimators (N2 and O3N2, respectively).
\label{fig:highz}}
\end{figure}

In Fig.~\ref{fig:highz}, we show the comparison between the above observational sample
at $z\sim 2$ and our models. We adopt the DL07 model for the dust optical properties.
The fiducial case with $\tau_\mathrm{SF}=5$ Gyr and $\eta_\mathrm{dense}=0.5$
underpredicts the observation data by a factor of $\sim 5$.
As shown in Section \ref{subsec:Z_nearby}, the MIR emission is stronger for a lower dense gas
fraction; thus, we also examine the case of $\eta_\mathrm{dense}=0.1$
(with $\tau_\mathrm{SF}=5$ Gyr).
As we observe in Fig.\ \ref{fig:highz}, the evolution with $\eta_\mathrm{dense}=0.1$ broadly
reproduces the observational data. The rising trend of the 7.7 $\micron$--TIR ratio is
interpreted as increasing abundances of PAHs and small grains.
The fact that the observational data are better explained by $\eta_\mathrm{dense}\simeq 0.1$
implies that the major part of PAH emission comes from the diffuse ISM.

The sample may be biased to actively star-forming galaxies, which have shorter
star formation time-scales.
Thus, we also show the case of a short $\tau_\mathrm{SF}=0.5$~Gyr
(with $\eta_\mathrm{dense}=0.5$) in Fig.\ \ref{fig:highz}.
As also shown in Fig.\ \ref{fig:clr_tau},
the increase of the MIR-to-TIR intensity ratio occurs at a higher metallicity if
the star formation time-scale is shorter. However, we observe in Fig.\ \ref{fig:highz} that
the metallicity level at which the intensity ratio increases is too high
to explain the observational data at $z\sim 2$.
Thus, a longer $\tau_\mathrm{SF}(\sim 5~\mathrm{Gyr})$ is more favoured than a short
starburst.
Since the age of the Universe is $\sim 3$~Gyr at $z\sim 2$, $\tau_\mathrm{SF}$
comparable to the cosmic age is favoured. This implies that the
galaxies sampled by \textit{Spitzer} at $z\sim 2$
are forming stars in a continuous way
over the cosmic age rather than only
in a recent short starburst. {However, this needs to be confirmed with further
modelling efforts including more realistic star formation histories and additional observational data
(e.g.\ more detailed SEDs).}

\citet{Schreiber:2018aa} have also shown that the relation between 8 $\micron$--TIR ratio
and metallicity holds for a sample of star-forming galaxies at $z\sim 0$--4. They estimated the metallicity from
stellar masses and star formation rates using
the fundamental relation observationally proposed by \citet{Mannucci:2010aa}.
They obtained a similar relation to that found by \citet{Galliano:2008aa} at $z\sim 0$, indicating that
the 8 $\micron$--TIR ratio is regulated by metallicity as predicted in this paper.
This also means that the redshift evolution of PAH emission could be predominantly driven by
metallicity evolution.

\subsection{Implication for spatially resolved PAH distribution}\label{subsec:resolved_PAH}

In our one-zone model, the PAH emission (or 8~$\micron$ emission) is
as strong as observed for nearby galaxies if the ISM is dominated by the diffuse phase
(i.e.\ $\eta_\mathrm{dense}=0.1$ in our model; Fig.\ \ref{fig:8_TIR}a).
However, we should note that the 24 $\micron$ emission is overpredicted
with $\eta_\mathrm{dense}=0.1$ (Fig.\ \ref{fig:24_TIR}a).
In reality, the emission from various dust components may originate from
different ISM phases. Indeed, \citet{Popescu:2011aa} constructed an SED model
for NGC 891 based
on an analytic galactic structure model and showed that the 8 $\micron$ emission is dominated
by the diffuse component while the 24 $\micron$ radiation is more associated with compact star-forming
regions. Since we use a one-zone model, we are not able to treat
the variety in the grain size distributions in different ISM phases.
It is still interesting to calculate the spatially resolved evolution of dust
including PAHs. There are some possible approaches of treating dust evolution in a spatially
resolved way. One is extension to a multi-component ISM model,
in which we treat the different grain size distributions in various  ISM phases. In this way,
we will be able to clarify how the different grain size distributions are superposed.
{\citet{Millan:2020aa} constructed a model that takes into account the
time variation of multi-component
ISM phases and dust abundance consistently;
however, they did not yet consider the different dust abundances in different ISM phases.}
Another viable way is to develop galaxy simulations that implement the evolution of
grain size distribution as already done by
\citet{McKinnon:2018aa} and \citet{Aoyama:2020aa}.
We could utilize these simulations to further include PAH evolution.
In particular, \citet{Aoyama:2020aa} showed that the grain size distributions
are different in different ISM phases.
Future spatially resolved simulations including PAH evolution
may serve to resolve the above tension between the favoured values of $\eta_\mathrm{dense}$
in the two MIR (8 $\micron$ and 24 $\micron$) bands.

There are some spatially resolved PAH emission
maps in nearby galaxies. For example, \citet{Chastenet:2019aa}
showed that the PAH abundance in the diffuse ISM of the Large Magellanic Cloud (LMC) is as high as
that in the Milky Way. This could be interpreted as an enhanced small-grain abundance by
accretion in the intermediate metallicity range, by a large aromatic fraction in the
diffuse ISM, or by shattering in the diffuse ISM.
The abundance of small grains contributing to the excess
emission at 70 $\micron$ is also large in the diffuse ISM of the LMC \citep{Bernard:2008aa}, supporting
efficient small-grain production in the diffuse medium.
\citet{Paradis:2009aa} showed that the spatial distributions of PAHs and small grains are different
in such a way that small grains are more associated with star-forming regions.
There are also some pieces of evidence of very different spatial distributions between
small grains contributing to the 24 $\micron$ emission and PAHs emitting at shorter
wavelengths for nearby galaxies \citep{Onaka:2018aa}.
Hydrodynamic simulations appropriately including dust evolution and hydrodynamic mixing will
address these spatially resolved features within the LMC or within a galaxy in general.

On the other hand, \citet{Chastenet:2019aa} indicated similar PAH abundances
in the diffuse neutral medium and in the molecular clouds,
which implies some mechanism to
homogenize the PAH abundance within a galaxy.
This may indicate that our model that assumes
different PAH formation/destruction rates between the diffuse and dense ISM with fixed
gas densities is too simplistic.
Furthermore, \citet{Relano:2016aa} and \citet{Chastenet:2019aa} showed that the PAH abundance is
lower in ionized regions \citep[see also][]{Lebouteiller:2011aa}, which could indicate
that the PAH abundance is affected by some physical processes related to gas ionization
(e.g.\ UV radiation).
H\,\textsc{ii} regions could also affect the evolution of small grains.
\citet{Relano:2016aa} found in M33 that the abundance of small grains is enhanced in
H\,\textsc{ii} regions. This indicates that some disruptive process of dust is associated with
H\,\textsc{ii} regions. Shattering can be induced by shocks \citep{Jones:1996aa}
or by efficient gyroresonance in ionized media \citep{Yan:2004aa,Hirashita:2010aa}, leading to
enhanced small grain production. Therefore, we should keep in mind that processes not
included in our model could be important in interpreting the detailed spatial distributions of
small grains and PAHs.

Another observational approach of obtaining spatially resolved information for nearby galaxies
is to use H\,\textsc{ii} regions, where the metallicity and the dust emission can be
measured simultaneously {(although we should keep in mind the above remarks for dust processing
specific for ionized
regions)}. \citet{Khramtsova:2013aa} investigated the PAH abundances in
individual H\,\textsc{ii} complexes of nearby star-forming galaxies. They found that the relation between
the PAH-to-dust ratio and the metallicity for individual H\,\textsc{ii} complexes is similar to that
in entire galaxies.
\citet{Khramtsova:2014aa} showed that the 8--24 $\micron$ ratio for
individual H\,\textsc{ii} complexes in nearby star-forming galaxies is similar to that
found for entire galaxies. These similarities between individual H\,\textsc{ii} complexes
and entire galaxies imply
that the evolution of the entire galaxy is still important when we interpret the
spatially resolved information on the PAH emission.
{The relation between the local and global PAH abundances could be theoretically}
addressed in future development
of spatially resolved modelling mentioned above.

\subsection{Necessity of further refinement for the PAH models}\label{subsec:enhance_PAH}

DL07 (based on \citealt{Weingartner:2001aa}) and J13 also derived
the grain size distributions that reproduce the Milky Way extinction curve and SED
based on their adopted dust optical properties.
They both required a lognormal-like excess in the size distribution of small aromatic grains.
Thus, the reason why we tend to underproduce the PAH emission (or 8 $\micron$ emission)
for the fiducial dense gas fraction ($\eta_\mathrm{dense}=0.5$) is that our grain
size distributions do not have such a lognormal excess.
If the excess of small aromatic grains is really required, we should consider some
mechanism that could enhance the PAH abundance.

There could be some PAH formation paths that are not included in our model.
Stellar PAH production in AGB stars \citep[e.g.][]{Latter:1991aa} could contribute to
the PAH enrichment in galaxies \citep[e.g.][]{Galliano:2008aa}.
However, few carbon-rich evolved stars have shown direct observational evidence for
PAH formation \citep[e.g.][]{Buss:1991aa}.
Theoretically, the stellar PAH yield is highly uncertain because it is sensitive to
the physical conditions adopted for stellar envelopes and winds
\citep{Frenklach:1989aa,Cherchneff:1992aa}.
Even if PAHs are supplied by stars, they may not be the major population of the interstellar
PAHs. In many models including the current model, stellar dust formation cannot explain
the dust abundance in nearby
galaxies \citep[e.g.][]{Dwek:1998aa,Hirashita:1999aa} and in some high-redshift galaxies
\citep[e.g.][]{Mancini:2015aa} because of the destruction in SN shocks. Thus, they considered
dust growth by accretion in the ISM to supplement the interstellar dust.
Since PAHs are also efficiently destroyed in SN shocks \citep{Micelotta:2010aa},
stellar sources are not likely to dominate the interstellar PAH abundance, especially
in evolved high-metallicity galaxies,
and interstellar PAH production is necessary to counterbalance the SN destruction.
Thus, we argue that stellar PAH formation is not a viable candidate for a significant PAH
formation mechanism.
There could be another PAH production mechanism through interstellar processing.
\citet{Hoang:2019aa} showed that radiative torque can induce rotational
disruption of grains by centrifugal force. This mechanism could act as an additional formation path of
small grains and PAHs. However, \citet{Hirashita:2020ac} showed that rotational disruption
still has a difficulty in efficiently producing grains as small as PAHs.
{Shattering could occur even in the dense ISM if the medium is ionized
\citep{Yan:2004aa,Hirashita:2010aa}. As mentioned in Section \ref{subsec:resolved_PAH},
this could explain the observed enhancement of small grains in ionized regions; however, the
PAH abundances tend to be suppressed in ionized regions observationally.
\citet{Chastenet:2019aa} also proposed formation of PAHs in the dense ISM, but the physical
mechanism is to be clarified.}
Therefore, it is not
yet clear if these extra formation paths of small grains (rotational disruption and shattering)
help to enhance the PAH
abundance significantly. Further investigations of possible PAH formation mechanisms are necessary.

Another possible improvement of the model is to treat the evolution
of the ISRF spectrum and that of the grain size distribution consistently. To do this,
we need to solve radiation transfer within the galaxy based on the dust extinction
calculated from the grain size distribution. It is also necessary to synthesize the
stellar spectrum. In a hard ISRF, which is observed in low-metallicity galaxies, small PAHs may be
destroyed (Section \ref{subsec:gsd}).
Although the origin of the hard radiation field in low metallicity galaxies
is not fully understood, it is interesting to make an effort of modelling the ISRF and
the PAH evolution consistently.

For further refinement, a detailed analysis using spectral data may also be useful
\citep[e.g.][]{Smith:2007aa,Hunt:2010aa}. However, since the spectral shape depends on the
detailed PAH properties such as the ionization state (DL07), further knowledge on how
the physical properties of PAHs are modified in different
environments is necessary. Although this kind of knowledge is difficult to obtain, the comparison with
spectroscopic data, which contain much more information than photometric data,
is an important step to further understand the dust and PAH evolution in galaxies.

\section{Conclusion}\label{sec:conclusion}

We examine how the evolution of grain size distribution affects the dust emission SEDs of galaxies.
We separate the grains into silicate and carbonaceous dust, and further
divide the latter into aromatic and non-aromatic species.
In our model, there are two main parameters for the evolution of grain size distribution:
the dense gas fraction ($\eta_\mathrm{dense}$) and
the star formation time-scale ($\tau_\mathrm{SF}$).
We synthesize the dust emission SED based on temperature distribution functions
under an ISRF intensity scaled with $U$ ($U=1$ corresponds to the
Milky Way ISRF in the solar neighbourhood).

We find that the SED shape is characterized by weak MIR emission
in the early phase of galaxy evolution because the grain abundance is dominated by large grains.
In the case of $\tau_\mathrm{SF}=5$~Gyr, which is roughly appropriate for the Milky Way-like galaxies,
the IR intensity grows rapidly by accretion (dust growth) around $t\sim 1$~Gyr.
Since accretion mainly increases the small-grain abundance, the MIR intensity grows more rapidly
than the FIR intensity. After this epoch, small grains are converted to large grains by coagulation;
in this phase, the FIR intensity grows slightly more rapidly than the MIR intensity.

The physical state of the ISM, parametrized by the dense gas
fraction $\eta_\mathrm{dense}$ in our model, also affects
the SED evolution, especially in the later epochs, when the dust evolution
is mainly driven by interstellar processing.
For the `fiducial' case with $\eta_\mathrm{dense}=0.5$, the observed
Milky Way SED is covered by the SEDs predicted for $t={}$a few--10 Gyr
in our model; however, there is a relative tendency that
the intensity at the wavelengths where PAH features are prominent is 
underpredicted. This can be resolved if we assume a
smaller fraction of the dense gas fraction (e.g.\ $\eta_\mathrm{dense}=0.1$).
If $\eta_\mathrm{dense}$ is small, the abundance of small grains is enhanced
and the aromatic fraction is high.

We also compare our results with the \textit{Spitzer} data of nearby galaxies.
The observed 70--160~$\micron$ ratios indicate that dust is broadly irradiated by
ISRFs of $U=1$--30. Some low-metallicity ($Z\sim 0.2$ Z$_{\sun}$) galaxies
are confirmed to have high $U\sim 100$.
The 24 $\micron$--TIR and 8 $\micron$--TIR ratios are insensitive to the ISRF intensity,
especially at high metallicity. Although the 24 $\micron$--TIR ratio in the fiducial dense gas fraction
($\eta_\mathrm{dense}=0.5$) is consistent with the
observational data at high metallicity, an enhanced
fraction of the diffuse ISM ($\eta_\mathrm{dense}\sim 0.1$) is required to reproduce
the observed 8 $\micron$--TIR ratios {at $Z\gtrsim 0.3$ Z$_{\sun}$}.
The different favoured values for $\eta_\mathrm{dense}$ between the 8 and 24 $\micron$ emissions
imply that they originate predominantly from different ISM phases.
A long star formation time-scale ($\tau_\mathrm{SF}\sim 50$ Gyr) is needed to explain the high
24 $\micron$--TIR and 8 $\micron$--TIR ratios at low metallicity in the nearby galaxy sample.
This is consistent with
the slow chemical enrichment of nearby low-metallicity galaxies.
{Some objects with very high 8 $\micron$--TIR ratios are hard to explain, so that it is
worth investigating an additional PAH production mechanism that is not included in our model.}

Finally, we adopt $z\sim 2$ galaxies \citep{Shivaei:2017aa} for further comparison.
A low dense gas fraction ($\eta_\mathrm{dense}\sim 0.1$) with a Milky Way-like star formation
time-scale ($\tau_\mathrm{SF}\sim 5$ Gyr) explains the rest-frame 7.7 $\micron$--TIR ratios. 
The positive correlation between the metallicity and the 7.7 $\micron$--TIR ratio is
interpreted by the increase of the PAH abundance by dust processing (shattering and accretion).
A short star formation time-scale appropriate for starburst galaxies
($\tau_\mathrm{SF}\sim 0.5$ Gyr) fails to explain the
observations since the increase of the PAH abudnance occurs at too high a metallicity.
This {may suggest} that the galaxies detected by the PAH band (24 $\micron$)
at $z\sim 2$ are forming stars
with a time-scale similar to nearby spiral galaxies (like the Milky Way).

It is worth emphasizing that the above results for PAHs (or emissions at $\lambda\sim 8~\micron$)
favour the ISM dominated by the diffuse ISM ($\eta_\mathrm{dense}\sim 0.1$), while
we needed a significant fraction of the dense ISM ($\eta_\mathrm{dense}\sim 0.5$)
to explain the extinction curves in HM20. The 24 $\micron$--TIR ratio also supports
$\eta_\mathrm{dense}\sim 0.5$. This tension may indicate that spatially
inhomogeneous dust evolution could be important, and that the emissions from different dust
components have different weights for the ISM phases. This issue could be {at least partially}
addressed if we
include our dust evolution model in hydrodynamic simulations.
{Possibilities of enhanced small-grain production and suppression of PAHs in
ionized regions as indicated in nearby galaxies also need further investigation and modelling.}

\section*{Acknowledgements}
 
{We are grateful to the anonymous referee for useful comments.}
HH thanks the Ministry of Science and Technology for support through grant
MOST 107-2923-M-001-003-MY3 and MOST 108-2112-M-001-007-MY3, and the Academia Sinica
for Investigator Award AS-IA-109-M02.
MSM acknowledges the support from the RFBR grant 18-52-52006.

\section*{Data availability}

{
The data underlying this article are available in Figshare at \url{https://doi.org/10.6084/m9.figshare.12917375.v1}.
}



\bibliographystyle{mnras}
\bibliography{/Users/hirashita/bibdata/hirashita}


\appendix


\bsp	
\label{lastpage}
\end{document}